\documentclass[12pt]{article}

\usepackage{amsmath}
\usepackage{graphicx}
\usepackage{enumerate}
\usepackage{natbib}
\usepackage{url} 
\usepackage{amsmath}
\usepackage{xcolor}

\usepackage{graphicx}
\usepackage{enumerate}
\usepackage{natbib}
\usepackage{caption}
\usepackage{subcaption}
\usepackage{float}
\usepackage{ amssymb }
\usepackage{pdflscape}

\usepackage{array}                
\usepackage{booktabs}             
\usepackage{multirow}             
\usepackage{colortbl}             
\usepackage[bottom]{footmisc}     

\usepackage{amsthm}
\usepackage{amsmath}
\usepackage{amssymb}
\usepackage{hyperref}

\usepackage[ruled,linesnumbered]{algorithm2e}

\newcommand{\blind}{1}

\def \tr{\mbox{tr}}

\def \Sigmab{\boldsymbol{\Sigma}}
\def \Kb{\boldsymbol{K}}
\def \Ub{\boldsymbol{U}}
\def \Xb{\boldsymbol{X}}

\newtheorem{lemma}{Lemma}

\newtheorem{theorem}{Theorem}

\begin{document}

\def\spacingset#1{\renewcommand{\baselinestretch}%
{#1}\small\normalsize} \spacingset{1}

\if1\blind
{
  \title{\bf Scalable Bayesian Structure Learning for Gaussian Graphical Models Using Marginal Pseudo-likelihood}
  \author{Reza Mohammadi, Marit Schoonhoven, Lucas Vogels and Ş. İlker Birbil \\
    Department of Business Analytics, Faculty of Economics and Business, \\ University of Amsterdam}
  \maketitle
} \fi

\begin{abstract} 
Bayesian methods for learning Gaussian graphical models offer a principled framework for quantifying model uncertainty and incorporating prior knowledge. However, their scalability is constrained by the computational cost of jointly exploring graph structures and precision matrices. To address this challenge, we perform inference directly on the graph by integrating out the precision matrix. We adopt a marginal pseudo-likelihood approach, eliminating the need to compute intractable normalizing constants and perform computationally intensive precision matrix sampling. Building on this framework, we develop continuous-time (birth–death) and discrete-time (reversible jump) Markov chain Monte Carlo (MCMC) algorithms that efficiently explore the posterior over graph space. We establish theoretical guarantees for posterior contraction, convergence, and graph selection consistency. The algorithms scale to large graph spaces, enabling parallel exploration for graphs with over 1,000 nodes, while providing uncertainty quantification and supporting flexible prior specification over the graph space. Extensive simulations show substantial computational gains over state-of-the-art Bayesian approaches without sacrificing graph recovery accuracy. Applications to human and mouse gene expression datasets demonstrate the ability of our approach to recover biologically meaningful structures and quantify uncertainty in complex networks. An implementation is available in the R package \textsf{BDgraph}.
\end{abstract}



\section{Introduction}
\label{sec:intro} 

Undirected graphical models \citep{lauritzen1996graphical, koller2009probabilistic} serve as fundamental tools for analyzing conditional dependencies among variables. A conditional dependency represents the association between variables conditional on the presence of other variables. These dependencies are naturally represented by graphs, where nodes correspond to random variables \citep{lauritzen1996graphical}, and the absence of an edge between two nodes signifies conditional independence \citep{rue2005gaussian}. The process of estimating this underlying graph structure is known as \textit{structure learning}.  

In this article, we consider Bayesian structure learning approaches for estimating Gaussian graphical models (GGMs), in contrast to frequentist techniques such as neighborhood selection \citep{meinshausen2006high} and the optimization of the likelihood function \citep{friedman2008sparse}. Bayesian methods offer key advantages, including the ability to quantify model uncertainty through posterior distributions and incorporate prior knowledge. However, their computational scalability often lags behind frequentist alternatives as the dimensionality of the problem increases.


The primary goal of Bayesian Structure Learning is to infer the underlying graph given the observed data. This is typically achieved by computing the posterior distribution of the graph conditional on the data. For GGMs, this requires the evaluation of complex integrals, which becomes increasingly challenging or even infeasible for large-scale graphs. Consequently, most Bayesian methods compute the joint posterior distribution of the graph and precision matrix. These methods have two primary bottlenecks in each Markov chain Monte Carlo (MCMC) iteration: (i) approximating intractable normalizing constants and (ii) iteratively updating the precision matrix. Consequently, full joint posterior exploration becomes computationally prohibitive for graphs exceeding 100 nodes in reversible jump and birth-death MCMC algorithms \citep{mohammadi2015bayesianStructure}, and 250 nodes in the spike-and-slab approach \citep{wang15}. To mitigate these limitations, \citet{mohammadi2023accelerating} introduced an MCMC-based method incorporating normalizing constant approximations, enabling scalability to a few hundred nodes. Similarly, \citet{van2022g} proposed a \(G\)-Wishart weighted proposal algorithm that leverages delayed acceptance MCMC and an informed proposal distribution to reduce the computational costs. These methods, despite their advances, remain computationally infeasible for modern applications involving thousands of variables.

Alternatively, several techniques have been proposed that bypass the full exploration of the graph space. They include methods based on multiple testing of marginal and conditional independence relationships \citep{williams2020bayesian, leday2019fast}. These methods are effective for large-scale problems but primarily focus on controlling the type I error rather than optimizing goodness-of-fit \citep{drton2007multiple}. Another approach avoids sampling over the graph space entirely by sampling solely from the posterior distribution of the precision matrix. Examples include block Gibbs samplers \citep{wang12, li2019graphical, sagar2024precision} and, recently, low-rank matrix decomposition methods \citep{chandra2024bayesian}. Although these approaches improve computational efficiency, they do not fully explore the posterior distribution of the graph space, restricting their ability to quantify model uncertainty. Moreover, approaches that focus on the precision matrix rather than the graph itself lack priors on the graphical structure and require additional steps to infer the underlying graph.

Another strategy to gain scalability is the approximation of the Gaussian likelihood. It has been applied to Bayesian structure learning for GGMs, most successfully by \citet{atchade2019quasi} and \citet{jalali2020b}. Both methods back their approximations with theoretical guarantees and push the boundaries of computational efficiency. Their algorithms still require sampling a precision matrix at every iteration, which poses challenges for scalability in high-dimensional settings. Instead, one could integrate out the precision matrix and target only the posterior on the graph space. The resulting marginal likelihood is not available in closed-form, but can be approximated using the pseudo-likelihood approximation by \citet{besag1975statistical}, resulting in the marginal pseudo-likelihood (MPL). Early studies \citep{pensar2016mpl, dobra2015} applied MPL to undirected graphical models with discrete variables. \citet{consonni2012objective, carvalho2009objective, stingo2015efficient} extended MPL to GGMs but initially restricted it to decomposable graphs. \citet{leppa2017learning} later adapted MPL to non-decomposable graphs via a score-based hill-climbing algorithm. However, this method only estimates the maximum a posteriori probability rather than fully characterizing posterior uncertainty. 

\textbf{Our Proposed Method and Key Contributions:}  
This article makes two main contributions to Bayesian structure learning in GGMs.  
First, we introduce a scalable framework capable of fully exploring the graph space.  
Second, we provide theoretical guarantees for consistency and convergence, which ensure reliable inference in large-scale graphical models.

We propose a scalable Bayesian framework for Gaussian graphical models by replacing the Gaussian likelihood with a marginal pseudo-likelihood (MPL) formulation. While the MPL approximation has been explored previously in Bayesian structure learning \citep{pensar2016mpl}, we extend its use to the design of scalable MCMC-based algorithms that operate directly in the graph space. Specifically, we develop two algorithms that combine the MPL approach with birth–death and reversible jump MCMC frameworks, enabling scalable and parallelizable exploration of large graph spaces and making inference feasible for graphs with more than 1,000 nodes. Our framework differs from the likelihood approximation methods of \citet{atchade2019quasi} and \citet{jalali2020b} in two important respects. First, their methods require sampling a precision matrix at every iteration, whereas our approach bypasses the precision matrix space entirely and samples solely over the graph space, leading to substantial gains in computational efficiency. Second, their methods approximate the full likelihood, namely the probability of the data given the precision matrix, while we approximate the marginal likelihood, namely the probability of the data given the graph.

Beyond scalability, we establish theoretical guarantees for our approach. We prove that the pseudo-posterior concentrates around the true posterior as the sample size increases. Moreover, Theorem \ref{thm:main} establishes the consistency of our algorithms, ensuring recovery of the true graph as both the sample size and the number of MCMC iterations grow. The theoretical guarantees we establish are general, applying not only to our proposed algorithms but to any MCMC procedure that targets the pseudo-posterior. In addition, our framework provides uncertainty quantification through edge inclusion probabilities and other graph characteristics, as illustrated in Section \ref{sec:application}. It can also incorporate prior knowledge about the graph structure, making the method adaptable to diverse applications.

\begin{figure}[!ht] 
\begin{center}
\includegraphics[width=4.5in]{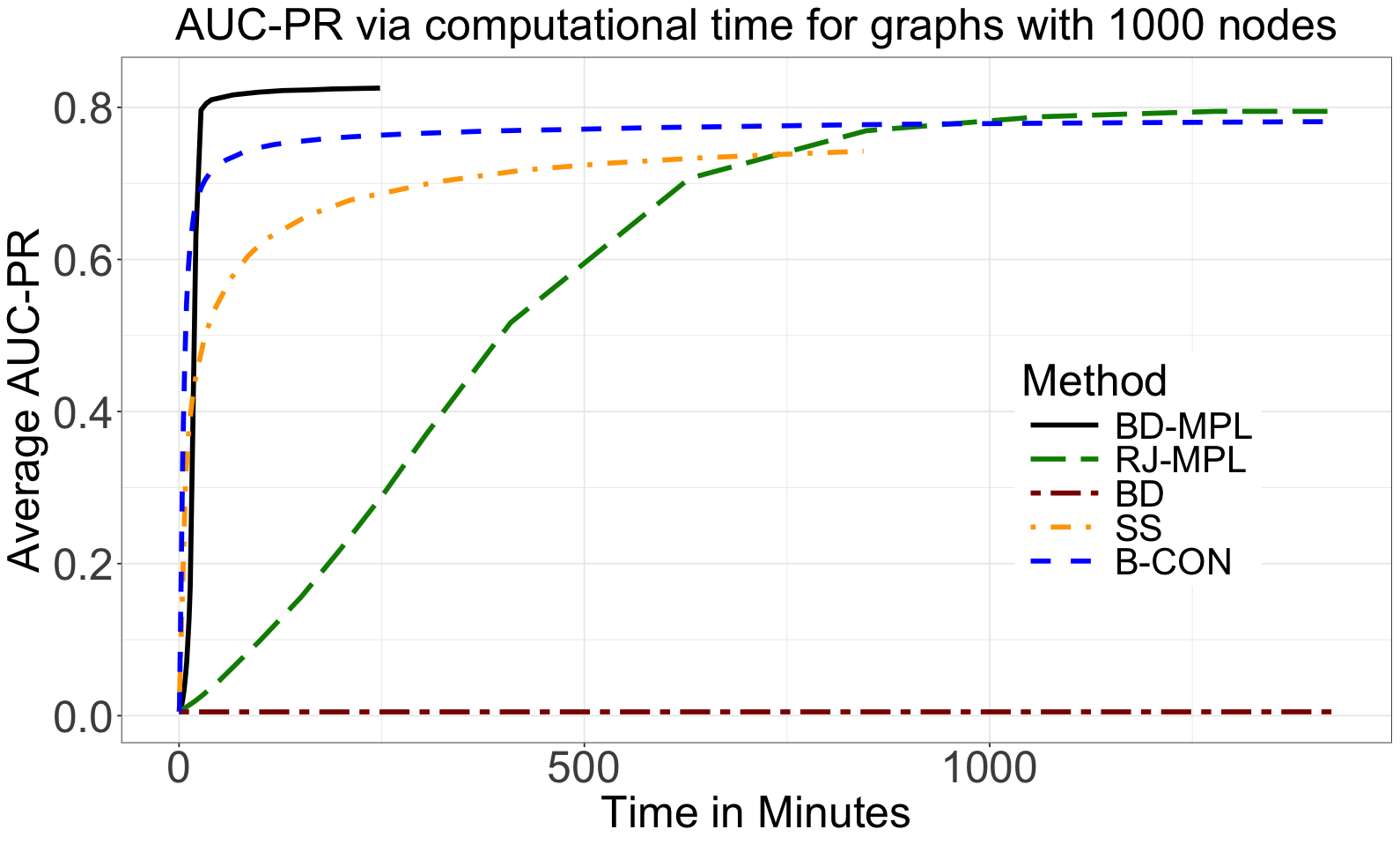}
\end{center}
\caption{\textit{
Average AUC-PR over running time for five algorithms applied to a simulated \textsf{Cluster} graph with 1,000 nodes, 0.5\% edge density, and 1,050 observations (see Section \ref{sec:simulation}), with 8 replications. BD-MPL and RJ-MPL are our proposed algorithms (Algorithms \ref{alg:BDMCMC} and \ref{alg:RJMCMC}, respectively). SS is the spike-and-slab method of \citet{wang15}, BD is the birth–death MCMC algorithm of \citet{mohammadi2023accelerating}, and B-CON is the method of \citet{jalali2020b}.
}
\label{fig:fig1}}
\end{figure}

Complementing our theoretical results, we present an extensive simulation study in Section \ref{sec:simulation} to assess the practical performance of our proposed algorithms and compare them with the state-of-the-art Bayesian structure learning methods for GGMs. To highlight improvements in computational efficiency and graph recovery accuracy, Figure \ref{fig:fig1} shows the convergence of the Area Under the Precision-Recall Curve (AUC-PR) over running time for our proposed algorithms (BD-MPL and RJ-MPL), alongside leading alternatives: the spike-and-slab (SS) method \citep{wang15}, the birth-death (BD) algorithm \citep{mohammadi2015bayesianStructure, mohammadi2023accelerating}, and the B-CONCORD (B-CON) method \citep{jalali2020b}. The simulation is based on a \textsf{Cluster} graph with 1,000 nodes, an edge density of 0.5\%, and 1050 observations. The results demonstrate the superior convergence speed of our BD-MPL algorithm, which achieves an AUC-PR above 0.8 in under 10 minutes. In comparison, the B-CON algorithm requires approximately 30 minutes to reach a moderate AUC-PR and fails to match BD-MPL’s performance even after one full day. The SS method also takes nearly a day to achieve a reasonable AUC-PR, but still lags in overall accuracy. The BD algorithm, by contrast, struggles considerably in this large-scale setting, remaining near an AUC-PR of 0.0 even after several days of computation.

The article is organized as follows. Section \ref{sec:BSL} introduces the fundamental concepts of Bayesian structure learning for GGMs. Section \ref{sec:BSL_MPL} presents the MPL approach and the two proposed MCMC-based algorithms for large-scale graph recovery. Section \ref{sec:theory} establishes the theoretical properties of the algorithms, including posterior contraction (Lemma \ref{lemma:contraction}), convergence (Lemma \ref{lemma:balance}), and graph selection consistency (Theorem \ref{thm:main}). Section \ref{sec:simulation} provides a comprehensive simulation study assessing the computational efficiency and accuracy of our methods compared to leading Bayesian approaches. In Section \ref{sec:application}, we demonstrate the versatility of our methods in uncertainty quantification through two real-world applications, showcasing their strengths on both medium- and large-scale datasets. Finally, we conclude with reflections and future research directions. Our implementation is available in the R package \textsf{BDgraph} \citep{BDgraph}. 

\section{Bayesian Structure Learning for GGMs}
\label{sec:BSL}

We denote an undirected graph as $G = (V, E)$, where $V$ is the set of $p$ nodes representing variables, and $E \subset \{(i,j) \mid 1 \leq i < j \leq p \}$ is the set of edges. An edge $(i,j) \in E$ indicates a connection between nodes $i$ and $j$, where each node corresponds to a distinct random variable, collectively forming a $p$-dimensional random vector. The observed data matrix is $\Xb = \left( \Xb^{(1)}, \ldots, \Xb^{(n)} \right)^T$ with dimensions $n \times p$, where each independent sample $\Xb^{(k)}$ ($k \in \{1, \ldots, n\}$) is a $p$-dimensional random vector. In GGMs, each $\Xb^{(k)}$ follows a multivariate Gaussian distribution $\mathcal{N}_p \left(\boldsymbol{0}, \Sigmab \right)$, where $\Sigmab$ is the covariance matrix and $\Kb = \Sigmab^{-1}$ is the precision matrix with elements $K_{ij}$. Two nodes $i$ and $j$ are conditionally independent if and only if $K_{ij} = 0$ \citep{lauritzen1996graphical}.  

In Bayesian structure learning, the goal is to estimate the posterior probability of a graph \( G \) given the data \( \Xb \)
\begin{equation}
\label{eq:p(g|x)}
P(G|\Xb) \propto P(G) P(\Xb|G),
\end{equation}  
where $P(G)$ is a prior distribution over the graph-space $\mathcal{G}_p$ of undirected graphs with $p$ nodes, and $P(\Xb|G)$ represents the marginal likelihood of $G$. A common specification for the prior $P(G)$ assumes independent edge inclusion probabilities $\beta_{ij} \in (0,1)$ for each edge $e = (i,j)$, enabling the incorporation of prior domain knowledge. This flexible formulation allows structural information to be encoded, for example, by assigning higher values of $\beta_{ij}$ to edges believed to exist and lower values otherwise. When all $\beta_{ij}$ are set to a common value $\beta \in (0,1)$, the prior simplifies to
\begin{equation}
\label{eq:p(g)}
P(G) \propto \beta^{|E|} (1-\beta)^{|\bar{E}|},
\end{equation}
where $|E|$ denotes the number of edges in $G$, and $|\bar{E}|$ is the number of absent edges. Smaller values of $\beta$ favor sparser graphs. When $\beta = 0.5$, the prior becomes uniform over the graph space. The hyperparameter $\beta$ directly controls the expected number of edges and encodes prior beliefs about graph sparsity. It is important to note that our Bayesian approach is not restricted to this prior form and can accommodate any prior distribution on $G$. For alternative graph prior specifications, we refer the reader to \cite{scutari2013prior, vandenboom2023}.

For the marginal likelihood of $G$, we have
\begin{equation}
\label{eq:p(x|g)}
P(\Xb|G) = \int_{\Kb} P(\Xb|G, \Kb)P(\Kb|G)d\Kb,
\end{equation}
where $P(\Kb|G)$ denotes the prior for $\Kb$ given $G$ and $P(\Xb|G, \Kb)$ is the likelihood function. 
A well-defined choice for the prior distribution of the precision matrix $\Kb$ is the $G$-Wishart distribution \citep{roverato2002hyper, letac2007whishart}, which serves as the conjugate prior for the multivariate Gaussian likelihood. The $G$-Wishart density is
\begin{equation}
\label{eq:gwish}
P(\Kb|G)=\frac{1}{I_G(b,\boldsymbol{D})}|\Kb|^{\frac{b-2}{2}} \exp\left\{\frac{-1}{2} \tr(\Kb\boldsymbol{D})\right\} \mathbf{1}\left(\Kb \in P_G\right),
\end{equation}
where $|\Kb|$ denotes the determinant of $\Kb$, $\tr(\boldsymbol{A})$ is the trace of a square matrix $\boldsymbol{A}$, $I_G(b,\boldsymbol{D})$ is the normalizing constant, and $P_G$ is the set of positive definite matrices $\Kb$ with $K_{ij}=0$ if $(i,j) \notin E$, 
and \(\mathbf{1}\left(\Kb \in P_G\right)\) is an indicator function that equals 1 if \(\Kb \in P_G\) and 0 otherwise. We denote this distribution with $W_G(b, \boldsymbol{D})$, where the symmetric positive definite matrix $\boldsymbol{D}$ and the scalar $b>2$ are the scale and the shape parameters of the $G$-Wishart distribution, respectively. 
Using the $G$-Wishart prior, \eqref{eq:p(x|g)} becomes
\begin{equation*}
P(\Xb|G)=(2\pi)^{-\frac{np}{2}}\frac{I_G(b+n, \boldsymbol{D}+\Ub)}{I_G(b,\boldsymbol{D})},
\end{equation*}
where $\Ub=\Xb^T\Xb$. Since this ratio of normalizing constants is intractable \citep{atay2005monte, mohammadi2023accelerating, wong2024new, wong2025conjecture, uhler2018exact}, most Bayesian methods circumvent it with MCMC algorithms that sample over the joint space of graphs and precision matrices. 

The joint posterior distribution of the graph $G$ and the precision matrix $\Kb$ is 
\begin{equation}
\label{eq:p(g,k|x)}
\begin{split}
P(G, \Kb|\Xb) & \propto P(\Xb|\Kb, G)P(\Kb|G)P(G) \\
&\propto P(G) \frac{1}{I_G(b,\boldsymbol{D})}|\Kb|^\frac{b+n-2}{2} \exp\left\{\frac{-1}{2} \tr\left(\Kb(\boldsymbol{D}+\Ub) \right)\right\}. 
\end{split}
\end{equation}
Computing this posterior distribution for all graphs $G \in \mathcal{G}_p$ is computationally infeasible for $p > 10$ due to the exponential number of possible graphs. Consequently, most Bayesian methods rely on MCMC-based algorithms, such as the reversible jump MCMC algorithm \citep{green1995reversible}, a discrete-time Markov chain approach \citep{dobra2011bayesian, dobra2011computational, cheng2012hierarchical, lenkoski2013direct, hinne2014efficient}. reversible jump MCMC explores the graph space by proposing to add or remove a single edge at each iteration, accepting the move with a probability that depends on the ratio of posterior probabilities (the conditional Bayes factor). However, reversible jump MCMC often suffers from low acceptance rates, requiring many iterations to converge. To mitigate this, \citet{mohammadi2015bayesianStructure} proposed a continuous-time MCMC algorithm, in which transitions between neighboring graphs, via edge additions or deletions, are modeled as independent Poisson processes.

A major computational bottleneck in these algorithms is evaluating the ratio of posterior probabilities, which requires computing expensive normalizing constants. Additionally, each new graph necessitates deriving a precision matrix by sampling from the \(G\)-Wishart distribution, further increasing computational overhead. Several improvements have been proposed to these challenges -- for a comprehensive review, see \cite{vogels2024bayesian} -- but efficient exploration of large graph space remains challenging.

\section{Bayesian Structure Learning with MPL}
\label{sec:BSL_MPL}


Recall that we aim to reduce the computational cost by sampling directly over the graph space instead of the joint space of graphs and precision matrices using MCMC-based search algorithms. To achieve this, we introduce two MCMC-based search algorithms that employ the MPL approach in conjunction with birth-death and reversible jump MCMC algorithms. In Section \ref{sec:MPL}, we illustrate how the MPL approach facilitates the derivation of Bayes factors for MCMC algorithms. The birth-death and reversible jump MCMC-based algorithms are described in Sections \ref{sec: BDMCMC} and \ref{sec: RJMCMC}, respectively.



\subsection{Marginal Pseudo-Likelihood}
\label{sec:MPL}

Bayesian structure learning in GGMs relies on the development of computationally efficient search algorithms, such as those outlined in Sections \ref{sec: BDMCMC} and \ref{sec: RJMCMC}. A key component of these algorithms, often termed neighborhood search algorithms, is the calculation of Bayes factors between pairs of neighboring graphs as
\begin{equation}
\label{eq:mpl_0}
\frac{P(G'|\Xb)}{P(G|\Xb)} = \frac{P(\Xb|G')P(G')}{P(\Xb|G)P(G)},
\end{equation}
where the graphs $G = (V, E)$ and \(G' = (V, E')\) differ by a single edge \(e = (i, j)\), such that \(G' = (V, E \cup e)\) or \(G' = (V, E \setminus e)\). To compute \(P(G|\Xb)\), we require the marginal likelihood \(P(\Xb|G)\) in \eqref{eq:p(x|g)}, which does not have a closed-form expression. Instead, we use the pseudo-likelihood \citep{besag1975statistical}, which approximates the marginal likelihood with a product of conditional likelihoods as
\begin{equation}
\label{eq:mpl}
P(\Xb|G) \approx \tilde{P}(\Xb|G) := \prod_{h=1}^{p}P(\Xb_h|\Xb_{nb(h)}, G),
\end{equation}
where $\Xb_h$ is a $n$-dimensional vector corresponding to node $h$, $nb(h)$ refers to the set of neighbors of node $h$ with respect to $G=(V, E)$ and $\Xb_{nb(h)}$ is the sub-matrix of $\Xb$ corresponding to the nodes that are in $nb(h)$. We then have
\begin{equation}
\label{eq:mpl_2}
\begin{split}
\frac{P(\Xb|G')}{P(\Xb|G)} & \approx \frac{\prod_{h=1}^{p}P(\Xb_h|\Xb_{nb(h)}, G')}{\prod_{h=1}^{p}P(\Xb_h|\Xb_{nb(h)}, G)} \\
& = \frac{P(\Xb_i|\Xb_{nb(i)}, G')P(\Xb_j|\Xb_{nb(j)}, G')}{P(\Xb_i|\Xb_{nb(i)}, G) P(\Xb_j|\Xb_{nb(j)}, G)},
\end{split}
\end{equation}
where the last step is based on the fact that the graphs $G$ and $G'$ differ by a single edge $e=(i,j)$. To finalize the approximation of the Bayes factor \eqref{eq:mpl_0}, one task remains: a closed-form expression of the terms $P(\Xb_h|\Xb_{nb(h)}, G)$ in equation \eqref{eq:mpl_2}. However, a closed-form expression is intractable due to the presence of the intractable normalizing constants of the G-Wishart prior \eqref{eq:gwish}. Instead, \citet{leppa2017learning} provide an approximation
\begin{equation}
\label{eq:mpl_3}
P\left(\Xb_h|\Xb_{nb(h)}, G \right) \approx \tilde{P}\left(\Xb_h|\Xb_{nb(h)}, G \right) := \pi^{-\frac{n-1}{2}} \frac{\Gamma \left (\frac{n+p_h}{2} \right)}{\Gamma \left (\frac{p_h+1}{2} \right)} n^{-\frac{2p_h+1}{2}} \left( \frac{|\Ub_{nb(h) \cup {h}}|}{|\Ub_{nb(h)}|} \right)^{-\frac{n-1}{2}},
\end{equation}
where $p_h$ is the size of the neighborhood $nb(h)$, and $\Ub_{\boldsymbol{A}}$ denotes the submatrix of $\Ub$ corresponding to the variables in the set $\boldsymbol{A}$. For \eqref{eq:mpl_3} to be well defined, the matrices $\Ub_{nb(h)}$ and $\Ub_{nb(h)\cup{h}}$ must be positive definite, which requires $n \ge p_h + 1$. Since this must hold for all $h = 1, \ldots, p$, the condition becomes $n \ge \max \{p_h + 1 : h = 1, \ldots, p\}$. This requirement is automatically satisfied when $n \ge p$, but it often also holds when $n < p$ due to the typical sparsity of precision matrices. The condition fails only if a node has more than $n-1$ neighbors. For example, with $p = 100$ and $n = 50$, the condition is violated only if a node is connected to 50 or more neighbors, in which case $p_h = 50$.

To clarify the approximation in \eqref{eq:mpl_3}, we turn to directed acyclic graphical models (DAGs). \citet{consonni2012objective} derived a closed-form expression for the marginal likelihood $P(\Xb \mid M)$ of a Gaussian DAG, $M$. Their approach build on the \citet[Theorem 2]{geiger2002parameter} that proved the problem reduces to computing the marginal likelihood $P(\Xb_{\boldsymbol{A}} \mid M_c)$, where $\Xb_{\boldsymbol{A}}$ is the submatrix of $\Xb$ corresponding to the variables in the set $\boldsymbol{A}$, and $M_c$ is a complete DAG. In analogy to \eqref{eq:p(x|g)}, this can be written as
\begin{equation}
\label{eq:DAG_mg}
P(\Xb_{\boldsymbol{A}} \mid M_c) = \int_{\Kb} P(\Xb_{\boldsymbol{A}} \mid M_c, \Kb ) P(\Kb \mid M_c) \, d\Kb.
\end{equation}
To evaluate $P(\Xb_{\boldsymbol{A}} \mid M_c)$, we must specify a prior $P(\Kb \mid M_c)$. This prior should be proper (meaning it integrates to one), objective (containing no subjective information), and ideally computationally convenient for evaluating \eqref{eq:DAG_mg}. \citet{consonni2012objective} show that the Wishart prior $W(p,\Ub/n)$ satisfies these requirements. This prior is an example of a fractional (or data-dependent) prior, as it incorporates a fraction of the observed data. Crucially, it allows for a closed-form expression of $P(\Xb_{\boldsymbol{A}} \mid M_c)$ analogous to \eqref{eq:mpl_3}, as shown in \citet[Equations 24 and 25]{consonni2012objective}. These results extend to decomposable undirected graphs \citep{consonni2012objective, carvalho2009objective}. In particular, \citet[Equation 29]{consonni2012objective} derived a closed-form expression identical to \eqref{eq:mpl_3}, which is exact for decomposable graphs. For non-decomposable graphs, the equality no longer holds, but \citet{leppa2017learning} demonstrated that the right-hand side of \eqref{eq:mpl_3} still provides a useful approximation.

Consequently, our approach to evaluate the Bayes factor \eqref{eq:mpl_2} involves two layers of approximation: the use of a pseudo-likelihood in \eqref{eq:mpl}, and the approximation of its local components via \eqref{eq:mpl_3}. Combining these two approximations gives the pseudo-posterior as
\begin{equation}
\label{eq:pseudo_posterior}
    P(G \mid \Xb) \approx \tilde{P}(G \mid \Xb) \propto P(G) \prod_{h=1}^{p}\tilde{P}(\Xb_h|\Xb_{nb(h)}, G).
\end{equation}
Sampling from this pseudo-posterior requires computing the Bayes factor in \eqref{eq:mpl_2}, which can be done using the closed-form expression in \eqref{eq:mpl_3}. This requires only four evaluations of \eqref{eq:mpl_3}, making the computation highly efficient for MCMC-based search algorithms. In the following sections, we introduce two such algorithms for general undirected GGMs, applicable to both decomposable and non-decomposable graphs. Section \ref{sec:theory} provides the theoretical justification for employing the MPL approximation within our structure learning framework.


\subsection{Birth-Death MCMC Algorithm}
\label{sec: BDMCMC}

The birth–death MCMC algorithm, based on a continuous-time Markov process \citep{preston1975}, was applied to GGMs by \citet{mohammadi2015bayesianStructure} for sampling from the joint posterior of graphs and precision matrices \eqref{eq:p(g,k|x)}. We propose a modified version that incorporates the MPL approximation, allowing sampling exclusively over the graph space $\mathcal{G}_p$ from the pseudo-posterior \eqref{eq:pseudo_posterior}. At iteration \(s \in \{1, \ldots, S\}\), the state of the Markov chain is a graph \(G^{(s)}\), which transitions to \(G^{(s+1)}\) by either adding (birth) or removing (death) a single edge. These birth and death events are modeled as independent Poisson processes, each occurring with rate \(R_e(G)\). If a birth of edge \(e = (i,j)\) occurs, the process moves to \(G^{+e}=(V, E \cup e)\); if a death of edge \(e\) occurs, it moves to \(G^{-e}=(V, E \setminus e)\). Since the events are governed by independent Poisson processes, the waiting time between consecutive events follows an exponential distribution with mean
\begin{equation}
\label{eq:waiting}
W(G)=\frac{1}{\sum_{e} R_e(G)},
\end{equation}
where the summation is over all $e \in\{(i,j)| 1 \leq i < j \leq p \}$. The associated birth/death probabilities are
\begin{equation}
\label{eq:prob bd}
P(\text{birth/death of edge } e) = R_e(G) W(G) \text{   for all } e \in \left\{(i,j)| 1 \leq i < j \leq p \right\}.
\end{equation}

The birth-death MCMC search algorithm converges to the target posterior distribution \(P(G|\Xb)\) in \eqref{eq:p(g|x)} by setting the birth/death rates to
\begin{equation}
\label{eq:rate}
R_e(G) = \min \left\{ \frac{P(G'|\Xb)}{P(G|\Xb)}, 1 \right\}, \quad \text{for each } e \in \{(i,j) \mid 1 \leq i < j \leq p \},
\end{equation}
where $G'$ is either $G^{+e}$ or $G^{-e}$. \citet[Theorem 5.1]{dobra2015} showed that these rates lead to convergence of the algorithm to the target posterior distribution in \eqref{eq:p(g|x)}; see also Lemma \ref{lemma:balance}. Their proof relies on the detailed balance condition, which is sufficient for convergence, though not strictly necessary \citep{cappe2003}. This birth-death algorithm, which searches exclusively over the graph space, is referred to as BD-MPL. Algorithm \ref{alg:BDMCMC} provides the pseudo-code for this algorithm. 

\begin{algorithm}[H]
\label{alg:BDMCMC} 
\caption{BD-MPL search algorithm }
 \KwIn{ Data $\Xb$ and an initial graph $G_0 = (V,E)$. }
 \KwOut{ Samples from the posterior distribution \eqref{eq:p(g|x)}. } 
	Calculate in parallel the MPL of each node given $G_0$ by \eqref{eq:mpl_3}\;
	Calculate in parallel all the rates for each edge given $G_0$ by \eqref{eq:rate}\;
 \For{$S$ iterations}{
  \For{ the rates that need to be re-evaluated}{
   Calculate in parallel the birth and death rates by \eqref{eq:rate}\;
   }
   Calculate the waiting time by \eqref{eq:waiting}\;
   Update the graph by the birth/death probabilities in \eqref{eq:prob bd}\;
   Update the MPL of the two nodes associated with the flipped edge.
 }
\end{algorithm}
Algorithm \ref{alg:BDMCMC} offers a significant computational advantage, particularly in determining birth and death rates, which are well-suited for parallel execution. By retaining the marginal pseudo-likelihood of the current graph nodes, recalculations are required only for the two nodes associated with the flipped edge. This is why all marginal pseudo-likelihoods are initially calculated outside the main loop in line 1. Similarly, the majority of the birth and death rates remain unchanged between iterations, justifying their initial calculation outside the main loop in line 2. By retaining rates across iterations, only a small fraction requires re-evaluation. For a graph with \(p\) nodes, only \(2p-3\) of the possible \(p(p-1)/2\) rates need to be updated. For example, for a graph with \(p=100\) nodes, the BD-MPL algorithm recalculates just 197 rates per iteration, compared to the 4950 rates that would otherwise need updating in the traditional birth-death MCMC algorithm. These computational optimizations have been implemented in Algorithm \ref{alg:BDMCMC}, which is coded in C++ and ported to R. This implementation is available in the R package \textsf{BDgraph} \citep{BDgraph} as the \textsf{bdgraph.mpl()} function.

The output of Algorithm \ref{alg:BDMCMC} consists of a set of \(S\) sampled graphs, \(\left\{ G^{(1)}, \ldots, G^{(S)} \right\}\), and their corresponding waiting times, \(\left\{W^{(1)}, \ldots, W^{(S)} \right\}\). These outputs facilitate various types of inference. Specifically, the expected value of any information function \(f: \mathcal{G}_p \to \mathbb{R}\) can be approximated using the Rao-Blackwellized estimator \citep{cappe2003} as
\begin{equation}
\label{eq:E_f}
E\left(f(G)|\Xb\right) \approx \frac{\sum_{s=1}^S W^{(s)} f(G^{(s)})}{\sum_{s=1}^S W^{(s)}},
\end{equation}
which is well-suited for uncertainty quantification, as it accounts for posterior uncertainty. By defining different information functions \(f(G)\), various posterior probabilities can be computed. For instance, setting \(f(G)\) to detect highly connected nodes (hubs) enables inference on hub probabilities, as demonstrated in Section \ref{sec:simulation}. Similarly, defining \(f(G)\) as \(1\) when an edge \(e\) is in \(G\) and \(0\) otherwise allows the estimation of posterior edge inclusion probabilities as
\begin{equation}
\label{eq:p_ij}
\hat{P}_e = \hat{P}_e(e \in G^* | \Xb) = \frac{\sum_{s=1}^S W^{(s)} \mathbf{1}\left(e \in G^{(s)}\right)}{\sum_{s=1}^S W^{(s)}}.
\end{equation}
Here, $G^*$ denotes the underlying graph that encodes the true, but unknown, conditional dependence structure. Within the Bayesian model averaging framework, these probabilities provide a valuable summary of the explored graph space, highlighting the relative importance of all edges. They are commonly used for graph selection based on a threshold \(0 < v < 1\), with a typical choice of \(v = 0.5\) leading to the estimated graph \(\hat{G} = (V, \hat{E})\), where \(\hat{E} = \left\{e = (i, j) \mid \hat{P}_e \geq 0.5\right\}\). For graph estimation, we recommend this median-probability approach, as suggested by \citet{barbieri2004optimal}, over selecting the graph with the highest posterior probability.


Note that Algorithm \eqref{alg:BDMCMC} relies on the local approximation in \eqref{eq:mpl_3}, which is exact for decomposable graphs. Therefore, the algorithm can be adapted specifically for use with decomposable graphs through two straightforward modifications. First, the initial graph $G^{(0)}$ must be decomposable (for example, the empty graph satisfies this condition). Second, before computing the birth and death rates (lines 4 and 5 of the algorithm), we must check whether the proposed graph $G'$ remains decomposable, accepting only moves that preserve decomposability.

\subsection{Reversible Jump MCMC Algorithm}
\label{sec: RJMCMC}

To sample from the pseudo-posterior \eqref{eq:pseudo_posterior}, we present an alternative to BD-MPL (Algorithm \ref{alg:BDMCMC}). Specifically, we integrate the MPL approach with the reversible jump MCMC algorithm \citep{green1995reversible}, a discrete-time method that explores the graph space by proposing edge additions or deletions. In each iteration, the reversible jump MCMC algorithm proposes a new graph $G'$ by adding or deleting an edge from the current graph $G$. The acceptance probability for the proposed move, which ensures that the Markov chain has the correct stationary distribution, is
\begin{equation}
\label{eq:alpha}
\alpha(G, G')=min \left\{ \frac{P(G'|\Xb)q(G'|G)}{P(G|\Xb)q(G|G')}, 1 \right\},
\end{equation}
where \(q(G'|G)\) is the proposal probability of transitioning from graph $G$ to graph $G'$. Clearly, $q(G'|G) = 0$ when $G'$ and $G$ are not neighbors. When $G'$ and $G$ are neighbors, we assume a uniform proposal distribution. For neighboring graphs, we adopt a uniform proposal distribution, setting $q(G'|G) = q(G|G') = 1/nb_{\text{max}}$, where $nb_{\text{max}} = p(p-1)/2$ is the total number of graphs differing from $G$ by one edge (for alternative proposals, see \citealt{dobra2011bayesian, van2022g}). With this setup, $\alpha(G, G')$ aligns with the birth-death rate as defined in \eqref{eq:rate}, highlighting the connection between the reversible jump MCMC and birth-death MCMC algorithms. 


Our proposed reversible jump algorithm is abbreviated to RJ-MPL and the pseudo-code for this algorithm is described in Algorithm \ref{alg:RJMCMC}. Similar to Algorithm \ref{alg:BDMCMC}, we implement this algorithm in C++ and ported it to R. It is available within the R package \textsf{BDgraph} \citep{BDgraph} as \textsf{bdgraph.mpl()} function.

\begin{algorithm}[H]
\label{alg:RJMCMC} 
\caption{RJ-MPL search algorithm }
 \KwIn{ Data $\Xb$ and an initial graph $G_0 = (V,E)$. }
 \KwOut{ Samples from the posterior distribution \eqref{eq:p(g|x)}. } 
	Calculate in parallel the MPL of each node given $G_0$ by \eqref{eq:mpl_3}\;
 \For{$S$ iterations}{
   Draw a proposal graph by selecting an edge to flip\;
   Calculate the acceptance probability by \eqref{eq:alpha} and update the graph\;
   Update the MPL for the pair of nodes associated with the flipped edge.
 }
\end{algorithm}

The output of Algorithm \ref{alg:RJMCMC} consists of a set of \(S\) sampled graphs, \(\left\{ G^{(1)}, \ldots, G^{(S)} \right\}\), representing the posterior graph space. These samples can be leveraged for various types of inference and model uncertainty quantification by applying \eqref{eq:E_f}, where \(W^{(s)} = 1\) for \(s \in \{1, \ldots, S\}\). For example, the sampled graphs can be used to calculate the estimated edge-inclusion probabilities (\(\hat{P}_e\)) using \eqref{eq:p_ij}, where \(W^{(s)} = 1\) for \(s \in \{1, \ldots, S\}\). Similar to the BD-MPL approach discussed in Section \ref{sec: BDMCMC}, \(\hat{P}_e\) provides a means to quantify model uncertainty and can also be employed for model selection.

\subsection{Precision Matrix Estimation}
\label{sec:presitionMatrix}

The BD-MPL and RJ-MPL algorithms (Algorithms \ref{alg:BDMCMC} and \ref{alg:RJMCMC}) are designed to handle large-scale problems by recovering the underlying graph structure from the data. In practical applications, it may also be necessary to estimate the precision matrix. Here, we present two approaches for estimating the precision matrix using the graph samples generated by the BD-MPL and RJ-MPL algorithms.

One approach is first to estimate the underlying graph structure using the edge inclusion probabilities \eqref{eq:p_ij} derived from the BD-MPL or RJ-MPL algorithms. Specifically, we can obtain the estimated graph \(\hat{G} = (V, \hat{E})\) where \(\hat{E} = \left\{e = (i, j) \mid \hat{P}_e \geq 0.5\right\}\), and \(\hat{P}_e\) represents the estimated edge-inclusion probabilities \eqref{eq:p_ij}. This estimated graph can then be used to sample from the precision matrix using the \(G\)-Wishart distribution \eqref{eq:gwish}, where \(\Kb|\hat{G} \sim W_{\hat{G}}(b+n, \boldsymbol{D} + \boldsymbol{U})\), representing the posterior distribution of the precision matrix. This step can be performed using the sampling algorithm developed by \citet{lenkoski2013direct}, which is implemented in the R package \textsf{BDgraph} \citep{BDgraph} as \textsf{rgwish()} function. The precision matrix can then be estimated as the mean of the sampled matrices. It is important to emphasize that the estimated precision matrix will always be positive definite, as the sampling algorithm by \citet{lenkoski2013direct} guarantees positive definiteness.

Another approach is to use the sampled graphs \(\left\{G^{(1)}, \ldots, G^{(S)}\right\}\) generated by the BD-MPL or RJ-MPL algorithms, and then sample the corresponding precision matrices \(\left\{\Kb^{(1)}, \ldots, \Kb^{(S)}\right\}\) from \(W_{G}(b+n, \boldsymbol{D} + \boldsymbol{U})\), similar to the first approach. Note that, this method may not be feasible for very large-scale graphs, as saving all the sampled graphs can lead to memory issues; see, for example, \citet[Appendix]{mohammadi2019bdgraph}.

\section{Theoretical Properties}
\label{sec:theory}


Here, we provide theoretical results that validate the use of the pseudo-posterior $\tilde{P}(G \mid \Xb)$ in \eqref{eq:pseudo_posterior} as an approximation to the true posterior $P(G \mid \Xb)$ in \eqref{eq:p(g|x)}. Specifically, Lemma \ref{lemma:contraction} establishes that the pseudo-posterior contracts around the true graph. Lemma \ref{lemma:balance} shows that both proposed algorithms converge to the target pseudo-posterior. Most importantly, Theorem \ref{thm:main} demonstrates that any MCMC algorithm targeting the pseudo-posterior, including RJ-MPL and BD-MPL, yields a consistent estimator of the graph. These results require only the assumption of Gaussianity and impose no restrictions on dimensionality, minimal signal strength, or hyperparameter choices.

We start with the following lemma, which establishes posterior contraction. It states that the posterior mass assigned to the true sparsity pattern (the true graph $G^*$) converges to one in probability as $n$ tends to infinity. Equivalently, the posterior probability of graphs $G$ that differ from the true graph $G^*$ approaches zero as $n \to \infty$. 

\begin{lemma}[Posterior Contraction]
\label{lemma:contraction}
Let $\Xb = \left( \Xb^{(1)}, \ldots, \Xb^{(n)} \right)^T$ be an $n \times p$ data matrix, where each independent observation $\Xb^{(k)}$, for $k \in \{1, \ldots, n\}$, is distributed as $\mathcal{N}_p \left(\boldsymbol{0}, (\Kb^*)^{-1} \right)$. Let $G^* = (V, E^*)$ denote the true underlying graph that encodes the conditional independence structure implied by $\Kb^*$. Then, as $n \to \infty$, the pseudo-posterior distribution $\tilde{P}(G|\Xb)$ defined in \eqref{eq:pseudo_posterior} satisfies
\begin{equation*}
\tilde{P}(G^*|\Xb) \to 1 \quad \text{in probability}.
\end{equation*}
\end{lemma}

\begin{proof}
Based on Equation \eqref{eq:pseudo_posterior} we have
\[
\tilde{P}(G^*|\Xb)  \propto P(G^*) \prod_{j=1}^{p} \tilde{P}(\Xb_j|\Xb_{nb^*(j)}, G^*),
\]
where \(nb^*(j)\) denotes the set of neighbors of node $j$ and these neighbors together uniquely define the true graph $G^*$. Now, considering an arbitrary graph \(G'\) (not equal to $G^*$) with corresponding neighbors \(nb'(j)\) of node \(j \in \{1, \ldots, p\}\), we have
\[
\frac{\tilde{P}(G^*|\Xb)}{\tilde{P}(G'|\Xb)} = \frac{P(G^*)}{P(G')} \times \prod_{j=1}^p \frac{\tilde{P}(\Xb_j|\Xb_{nb^*(j)}, G^*)}{\tilde{P}(\Xb_j|\Xb_{nb'(j)}, G')}. 
\]
For all \(j \in \{1, \ldots, p\}\), as \(n \to \infty\), we can derive
\[
\log \left( \frac{\tilde{P}(\Xb_j|\Xb_{nb^*(j)}, G^*)}{\tilde{P}(\Xb_j|\Xb_{nb'(j)}, G')} \right) \to \infty
\]
in probability; this result follows directly from \citet[Theorem 2, Lemma 1, and 2]{leppa2017learning}. Essentially, this means that by using pseudo-posterior \eqref{eq:pseudo_posterior}, the true neighbors \(nb^*(j)\) are preferred over any other set \(nb'(j)\) as the number of observations \(n\) approaches infinity. Considering this, as \(n \to \infty\),
\[
\frac{\tilde{P}(G^*|\Xb)}{\tilde{P}(G'|\Xb)} \to \infty,
\]
in probability. The convergence then follows.
\end{proof}

The following lemma establishes that, as the number of MCMC iterations $S$ tends to infinity, the BD-MPL and RJ-MPL sampling algorithms converge to the pseudo-posterior distribution $\tilde{P}(G|\Xb)$. 
\begin{lemma} [Convergence]  
\label{lemma:balance}  
Let \(\left\{G^{(1)}, \ldots, G^{(S)}\right\}\) denote the Markov chain generated by either the BD-MPL or RJ-MPL algorithm. As the number of iterations \(S\) approaches infinity, the Markov chain converges to the target pseudo-posterior distribution \(\tilde{P}(G|\Xb)\) defined in \eqref{eq:pseudo_posterior}. Furthermore, for any function \( f: \mathcal{G}_p \to \mathbb{R} \), we have 
\begin{equation}  
\label{eq:E_f_equal}  
E\left(f(G)|\Xb\right) = \lim_{S \to \infty} \frac{\sum_{s=1}^S W^{(s)} f(G^{(s)})}{\sum_{s=1}^S W^{(s)}},  
\end{equation}  
where \(\left\{W^{(1)}, \ldots, W^{(S)} \right\}\) are the waiting times in the BD-MPL algorithm. For the RJ-MPL algorithm, these waiting times are equal to one.
\end{lemma}  

\begin{proof}  
Convergence requires three conditions: irreducibility, aperiodicity, and the balance condition, ensuring that the Markov chain has a well-defined stationary distribution \citep{stationary_distr_exists}. Both BD-MPL and RJ-MPL naturally satisfy irreducibility and aperiodicity. While detailed balance is a sufficient condition but not necessary for general balance, it is commonly imposed in practical sampler design \citep{green1995reversible}. For Algorithm \ref{alg:BDMCMC}, detailed balance holds by construction, as ensured by the birth-death rates defined in \eqref{eq:rate}; see \citet[Theorem 5.1]{dobra2015} for further details. Similarly, Algorithm \ref{alg:RJMCMC} satisfies these conditions through the acceptance probability defined in \eqref{eq:alpha} \citep{green1995reversible}. Since both BD-MPL and RJ-MPL converge to the target posterior distribution, equation \eqref{eq:E_f_equal} follows directly from the Rao-Blackwellized estimator \citep[Section 2.5]{cappe2003}.
\end{proof}  

An immediate consequence of Lemma \ref{lemma:balance} is the convergence of the estimated edge-inclusion probability \(\hat{P}_e\), as defined in \eqref{eq:p_ij}, to the true pseudo-posterior edge-inclusion probability \( P_e \). Specifically, for all \( e \in \left\{(i,j) \mid 1 \leq i < j \leq p \right\} \), we have  
\begin{equation*}
\lim_{S \to \infty} \hat{P}_e = P_e = \sum_{G \in \mathcal{G}_p} \mathbf{1}\left(e \in G\right) \tilde{P}(G|\Xb).
\end{equation*}  

The following theorem presents the main result. It involves the graph $\hat{G}$ obtained by thresholding the edge inclusion probabilities \eqref{eq:p_ij} at some threshold $0 < v < 1$, so that $\hat{G} = (V, \hat{E})$ with $\hat{E} = \{e = (i, j) \mid \hat{P}_e \geq v\}$. We prove that, for any fixed threshold $v$, the estimated graph $\hat{G}$ converges in probability to the true graph $G^*$ as both the number of observations and the number of MCMC iterations tend to infinity. In other words, the edge inclusion probabilities asymptotically converge to either 0 or 1, accurately reflecting the true edge structure of $G^*$. 

\begin{theorem}[Selection Consistency]
\label{thm:main}
Let $\hat{P}_e$ denote the estimated edge inclusion probabilities from Equation \eqref{eq:p_ij}, obtained via an MCMC sampling algorithm (such as BD-MPL or RJ-MPL) with $S$ iterations targeting the pseudo-posterior distribution \eqref{eq:pseudo_posterior}. Define the estimated graph $\hat{G} = (V, \hat{E})$, where an edge $e = (i,j)$ is included in $\hat{E}$ if $\hat{P}_e \geq v$ for some fixed threshold $0 < v < 1$. Then, as the number of observations $n \to \infty$ and the number of MCMC iterations $S \to \infty$,  
\begin{equation*}
P(\hat{G} = G^*) \;\to\; 1 \quad \text{in probability}.
\end{equation*}
\end{theorem}

\begin{proof}
Due to Lemma \ref{lemma:balance}, for all $e \in\{(i,j)| 1 \leq i < j \leq p \}$, we have
\begin{equation*}
\lim_{S \to \infty} \hat{P}_e = P_e = \sum_{G \in \mathcal{G}_p} \mathbf{1}\left(e \in G\right)\tilde{P}(G|\Xb). 
\end{equation*}
Considering $P_e \geq \tilde{P}(G^*|\Xb)$ for all $e \in G^*$ and $P_e \leq 1-\tilde{P}(G^*|\Xb)$ for all $e \not \in G^*$, we have
\begin{align*}
P(\hat{G}=G^*) & = P \left( P_e \geq v \quad \forall e \in G^* \text{ and } P_e < v \quad \forall e \not \in G^* \right) \\
& \geq P \left( \tilde{P}(G^*|\Xb) \geq v \text{ and } 1 - \tilde{P}(G^*|\Xb) < v \right) \\
& = P \left( \tilde{P}(G^*|\Xb) \geq max(v,1-v) \right).
\end{align*}
The result follows from Lemma \ref{lemma:contraction}.
\end{proof}

We note that, in connection with the theoretical results presented in this section, \citet[Theorem 1]{jalali2020b} also established posterior contraction and selection consistency for their B-CON method, which combines a spike-and-slab prior on the precision matrix with a generalized likelihood approximation to the Gaussian likelihood. Three key differences distinguish their results from ours. First, their framework requires structural conditions on the true precision matrix, including a minimal signal size and bounded eigenvalues \citep[Assumptions 3–4]{jalali2020b}, whereas our approach imposes no such constraints. Second, their results are tied to the specific graph prior in \eqref{eq:p(g)} with a fixed value of $\beta$ \citep[Assumption 5]{jalali2020b}, while ours hold for any prior on the graph. Third, our results assume Gaussian data, whereas B-CON requires only that the data-generating distribution has sub-Gaussian tails \citep[Assumption 2]{jalali2020b}.


Theorem \ref{thm:main} shows that an MCMC algorithm targeting the pseudo-posterior recovers the true graph as the sample size and number of iterations grow. We next evaluate its finite-sample performance through simulations against state-of-the-art Bayesian structure learning methods.

\section{Simulation Study}
\label{sec:simulation}


We assess the accuracy and computational efficiency of BD-MPL (Algorithm \ref{alg:BDMCMC}) and RJ-MPL (Algorithm \ref{alg:RJMCMC}) with three state-of-the-art Bayesian approaches\footnote{For comparisons with frequentist methods such as the graphical lasso \citep{friedman2008sparse}, we refer the reader to \citet{vogels2024bayesian}.}. The first is the birth-death MCMC algorithm (BD) introduced by \citet{mohammadi2015bayesianStructure} and \citet{mohammadi2023accelerating}. The second method, SS, is established by \citet{wang15}, and it employs a block Gibbs sampler based on the spike-and-slab prior distribution. The third is the B-CON approach developed by \citet{jalali2020b}, which uses a generalized likelihood function together with a spike-and-slab prior distribution.  

The simulation study covers small ($p=10$), medium ($p=100$), and large-scale ($p=1000$) graphs, considering three structural types. The first type, \textsf{Random}, consists of graphs in which edges are randomly sampled without replacement. The second type, \textsf{Cluster}, contains either two clusters (for $p \in {10,100}$) or eight clusters (for $p=1000$), with each cluster following the structure of a \textsf{Random} graph. The third type, \textsf{Scale-free}, comprises spanning trees generated using the B-A algorithm of \citet{albert2002}.
For both \textsf{Random} and \textsf{Cluster} graphs, we examine `\textit{sparse}' and `\textit{dense}' variants. To accurately represent these graphs, we determine the number of edges \(n_e\) using \(\max(ap, bp(p-1)/2)\), with \(a = 0.5\) and $b = 0.5\%$ for sparse graphs, and \(a = 2\) and \(b = 5\%\) for dense graphs. The edge densities of all the simulated graph types are reported in the Supplementary Material. For the number of observations \(n\), we follow a logarithmic relation between \(n\) and \(p\) as suggested by posterior contraction rates \cite[Theorem 4.6]{sagar2024precision}. We select \(n = 20 \log(p)\) for `few' observations and \(n = 350 \log(p)\) for `many' observations. An exception is made for \(p = 1000\), where \(n = 20 \log(p) = 60\) was too low for all algorithms to provide meaningful results, so we chose \(n = 400\).

In each simulated graph \(G\), the precision matrix \(\Kb\) was generated from the \(G\)-Wishart distribution \(W_G(3, \boldsymbol{I}_p)\). For \(p \in \{10, 100\}\), we generated 16 graphs along with their corresponding precision matrices, while for \(p = 1000\), we obtained 8 such pairs. Subsequently, for each pair \(G\) and \(\Kb\), we sampled \(n\) data points from the \(p\)-dimensional Gaussian distribution \(\mathcal{N}_p(\boldsymbol{0}, \Sigmab)\) with mean zero and covariance matrix \(\Sigmab = \Kb^{-1}\). The data was generated using the \textsf{bdgraph.sim()} function from the R package \textsf{BDgraph}. 

To ensure fair comparisons in computing time, all five algorithms were coded in C++ and then ported to R, using the same routines wherever feasible. Following the approach of \citet{wang15}, the hyperparameters for the SS method were chosen as $\epsilon = 0.02$, $\upsilon = 2$, and $\lambda = 1$. Each MCMC run started with an empty graph. For the prior on the graph, we use the distribution \eqref{eq:p(g)} with prior density $\beta$. In real-life cases, one can use domain knowledge to select an appropriate value for $\beta$. In a simulation study, however, such prior knowledge is not available. It is therefore common to set $\beta$ to either the uninformative value $\beta=0.5$ \citep{Gan2018} or to a function that decreases with $p$, for example $\beta=2/(p-1)$ \citep{van2022g}. Here, we opt for a middle ground and set $\beta=0.2$. For the B-CON method, we used $\beta = 0.5$ and initialized with a full graph, consistent with the authors’ original setup, since their provided code \citep{jalali2020b} does not allow specifying either the initial graph or $\beta$. The BD-MPL and RJ-MPL were implemented using the \textsf{BDgraph} R package \citep{BDgraph}, and the SS method utilized the \textsf{ssgraph} R package \citep{mohammadi2022ssgraph}. 

\begin{figure}[!ht]  
\centering  
\includegraphics[width=1\linewidth]{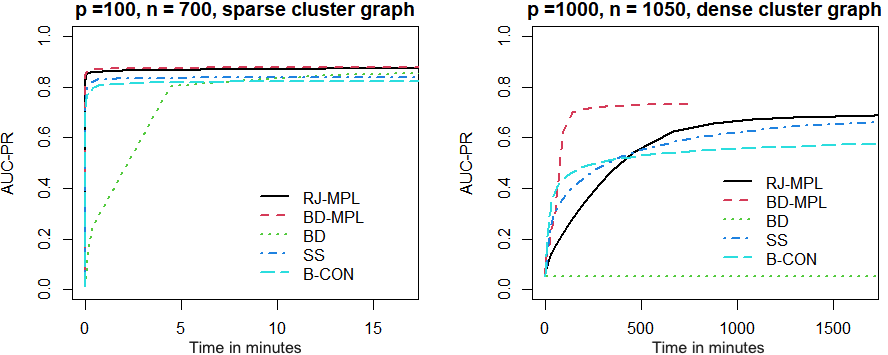}  
\caption{\textit{  
The convergence of AUC-PR scores over time for all algorithms (RJ-MPL, BD-MPL, BD, SS, B-CON). The left plot represents the instance with \(p = 100\), \(n = 700\) for the sparse \textsf{Cluster} graph, while the right plot corresponds to \(p = 1000\), \(n = 1050\) for the dense \textsf{Cluster} graph.  
}}  
\label{fig:sim_convergence_AUC-PR}  
\end{figure}

The methods are evaluated based on graph recovery accuracy and computational cost. To assess accuracy, we compute edge inclusion probabilities from the MCMC-sampled graphs \(\left\{ G^{(1)}, \ldots, G^{(S)}\right\}\) using \eqref{eq:p_ij}. Five accuracy metrics are considered, with the Area Under the Precision-Recall Curve (AUC-PR) \citep{davis2006roc-pr} highlighted here. The remaining metrics are provided in the Supplementary Material. AUC-PR is particularly useful for evaluating performance on imbalanced datasets, such as sparse graphs. The PR curve plots precision (the proportion of true positive edges among all predicted edges) against recall (the proportion of actual edges correctly identified) at various thresholds. Figure \ref{fig:sim_convergence_AUC-PR} presents AUC-PR scores over time for two scenarios: the left plot shows results for medium-scale graphs (\(p = 100\), \(n = 700\), \textit{sparse} \textsf{Cluster} graph), while the right plot corresponds to large-scale graphs (\(p = 1000\), \(n = 1050\), \textit{dense} \textsf{Cluster} graph). Additional instances are provided in the Supplementary Material. For the medium-scale problem, all methods except BD converge rapidly. In the large-scale setting, BD-MPL converges significantly faster in graph recovery precision compared to the other methods.  

To compare algorithms, we require a measure of computational cost. Ideally, this would capture the time needed for an MCMC chain to meet a formal convergence criterion. However, defining such a criterion for large-scale, MCMC algorithms is challenging. Standard diagnostics such as the Gelman–Rubin statistic \citep{Gelman1992}, effective sample size, or trace plots (Figure \ref{fig:trace_plot}) require storing entire chains (or multiple chains), which can be prohibitively memory-intensive at scale. We therefore adopt a more practical approach, stopping the chains when a relevant performance metric --- here the AUC-PR --- shows no meaningful improvement. Following \citet{vogels2024bayesian}, we define computational cost as the time required for the AUC-PR to come within 0.01 of its final iteration value. While this measure is inherently subjective, it serves as a useful indicator of the runtime needed for an algorithm’s output quality to stabilize rather than a formal proof of convergence. Table \ref{table:time-AUC-PR} summarizes the results. For $p = 10$, all algorithms converge in under a minute with similar costs. At $p = 100$, RJ-MPL, BD-MPL, SS, and B-CON converge within seconds to minutes, whereas BD requires several hours. For $p = 1000$, BD fails to converge within five days, while BD-MPL, RJ-MPL, SS, and B-CON converge within hours to days, with BD-MPL running up to ten times faster than the others.
\begin{table}[!ht]
\centering
\begin{tabular}{c @{\hspace{0.7em}} c  c c @{\hspace{2em}} c c c @{\hspace{1.7em}} c @{\hspace{0.7em}} c}
\arrayrulecolor{blue}\toprule
$p$   & Graph  & Density & $n$ &  RJ-MPL & BD-MPL & BD & SS & B-CON  \\ 
\hline 
\multirow{10}{*}{1000}&\multirow{4}{*}{\textsf{Random}}& Sparse  &  400  &2494  & \textbf{41}  & -&531  & 972   \\ 
                    &                         & Sparse  &  1050  &1998  & \textbf{96}  & -&543  & 418   \\ 
                    &                         & Dense  &  400  &2706  & \textbf{350}  & -&3012  & 3731   \\ 
                    &                         & Dense  &  1050  &2434  & \textbf{1065}  & -&3297  & 2099   \\ 
\cmidrule{2-9} 
                    &\multirow{4}{*}{\textsf{Cluster}} & Sparse  &  400  &2212  & \textbf{33}  & -&436  & 975   \\ 
                    &                         & Sparse  &  1050  &1794  & \textbf{74}  & -&489  & 615   \\ 
                    &                         & Dense  &  400  &2509  & \textbf{283}  & -&2386  & 3681   \\ 
                    &                         & Dense  &  1050  &3003  & \textbf{396}  & -&2654  & 1711   \\ 
\cmidrule{2-9} 
                 &\multirow{2}{*}{\textsf{Scale-free}} & Sparse  &  400  &4104  & \textbf{43}  & -&339  & 1808   \\ 
                    &                         & Sparse  &  1050  &4053  & \textbf{144}  & -&1281  & 1508  \\ 
\cmidrule{1-9} 
\multirow{10}{*}{100}&\multirow{4}{*}{\textsf{Random}} & Sparse  &  40  &6  & 2  & 103  & \textbf{0}  & 7   \\ 
                    &                         & Sparse  &  700  &\textbf{0}  & \textbf{0}  & 34  & 2  & 3   \\ 
                    &                         & Dense  &  40  &2  & \textbf{0}  & 101  & \textbf{0}  & 5   \\ 
                    &                         & Dense  &  700  &2  & \textbf{0}  & 72  & 1  & 3 \\ 
\cmidrule{2-9} 
                    &\multirow{4}{*}{\textsf{Cluster}} &  sparse  &  40  &6  & 3  & 86  & \textbf{1}  & 6   \\ 
                    &                         & Sparse  &  700  &1  & \textbf{1}  & 45  & \textbf{1}  & 4   \\ 
                    &                         & Dense  &  40  &\textbf{0}  & \textbf{0}  & 115  & \textbf{0}  & 4   \\ 
                    &                         & Dense  &  700  &1  & \textbf{0}  & 78  & 1  & 1   \\ 
\cmidrule{2-9} 
                &\multirow{2}{*}{\textsf{Scale-free}}  & Sparse  &  40  &2  & 1  & 103  & \textbf{0}  & 5   \\ 
                    &                         & Sparse  &  700  &2  & \textbf{1}  & 67  & \textbf{1}  & 4   \\ 
\bottomrule
\end{tabular}
\caption{\textit{Computational cost (\(T\)) in minutes until AUC-PR convergence for various instances. \(T\) represents the average time until AUC-PR convergence, based on 16 replications for \(p \in \{10, 100\}\) and 8 replications for \(p = 1000\). The table excludes the \(p=10\) case since the computational time for all algorithms was less than one minute. A ``-" indicates that an algorithm did not converge within five days. For each setting, the best-performing algorithm is highlighted in \textbf{bold}.}}
\label{table:time-AUC-PR}
\end{table}

Table \ref{table:AUC_PR} reports the AUC-PR scores for all methods. For small ($p = 10$) and medium ($p = 100$) graphs, performances are comparable, with BD showing slightly higher values. For large graphs ($p = 1000$), differences are pronounced: BD yields near-zero AUC-PR, reflecting its inability to handle large-scale problems within a reasonable runtime (five days in this study). In contrast, RJ-MPL and BD-MPL achieve the highest AUC-PR, followed by SS, B-CON, and BD. For example, in the dense \textsf{Cluster} graph with $p = 1000$ and $n = 1050$, BD-MPL reaches the highest AUC-PR (over 0.7) in under two hours, whereas SS and B-CON require more than 27 hours, and BD fails to converge within five days. Together with Table \ref{table:time-AUC-PR}, these results indicate that BD-MPL delivers the best overall performance for $p = 1000$, combining both computational cost and accuracy.
\begin{table}[!ht]
\centering
\begin{tabular}{c @{\hspace{0.7em}} c @{\hspace{0.7em}} c c @{\hspace{1.6em}} c c c @{\hspace{1.7em}} c @{\hspace{0.7em}} c}
\arrayrulecolor{blue}\toprule
$p$   & Graph  & Density & $n$ &  RJ-MPL & BD-MPL & BD & SS & B-CON  \\ 
\hline 
\multirow{10}{*}{1000}&\multirow{4}{*}{\textsf{Random}}& Sparse  &  400  &0.67  & \textbf{0.70}  & 0.01  & 0.66  & 0.66  \\ 
                    &                         & Sparse  &  1050  &0.79  & \textbf{0.81}  & 0.00  & 0.72  & 0.79   \\ 
                    &                         & Dense  &  400  &0.39  & \textbf{0.43}  & 0.05  & 0.42  & 0.31   \\ 
                    &                         & Dense  &  1050  &0.56  & \textbf{0.61}  & 0.05  & 0.56  & 0.50   \\ 
\cmidrule{2-9} 
                    &\multirow{4}{*}{\textsf{Cluster}} & Sparse  &  400  &0.70  & \textbf{0.72}  & 0.01  & 0.67  & 0.68   \\ 
                    &                         & Sparse  &  1050  &0.81  & \textbf{0.83}  & 0.01  & 0.74  & 0.78   \\ 
                    &                         & Dense  &  400  &0.55  & \textbf{0.59}  & 0.05  & 0.55  & 0.48   \\ 
                    &                         & Dense  &  1050  &0.71  & \textbf{0.74}  & 0.05  & 0.7  & 0.58   \\ 
\cmidrule{2-9} 
                 &\multirow{2}{*}{\textsf{Scale-free}} & Sparse  &  400  &0.66  & \textbf{0.68}  & 0.00  & \textbf{0.68}  & 0.62   \\ 
                    &                         & Sparse  &  1050  &0.78  & \textbf{0.8}  & 0.00  & 0.76  & 0.72   \\ 
\cmidrule{1-9} 
\multirow{10}{*}{100}&\multirow{4}{*}{\textsf{Random}} & Sparse  &  40  &0.50  & 0.50  & 0.54  & \textbf{0.55}  & 0.50   \\ 
                    &                         & Sparse  &  700  &\textbf{0.89}  & \textbf{0.89}  & \textbf{0.89}  & 0.84  & 0.86   \\ 
                    &                         & Dense  &  40  &0.37  & 0.37  & 0.40  & \textbf{0.43}  & 0.38   \\ 
                    &                         & Dense  &  700  &\textbf{0.86}  & \textbf{0.86}  & \textbf{0.86}  & 0.79  & 0.82   \\ 
\cmidrule{2-9} 
                    &\multirow{4}{*}{\textsf{Cluster}} & Sparse  &  40  &0.49  & 0.49  & 0.52  & \textbf{0.53}  & 0.49   \\ 
                    &                         & Sparse  &  700  &0.87  & \textbf{0.88}  & \textbf{0.88}  & 0.84  & 0.83   \\ 
                    &                         & Dense  &  40  &0.40  & 0.39  & 0.43  & \textbf{0.45}  & 0.41   \\ 
                    &                         & Dense  &  700  &\textbf{0.87}  & \textbf{0.87}  & \textbf{0.87}  & 0.80  & 0.83   \\ 
\cmidrule{2-9} 
                 &\multirow{2}{*}{\textsf{Scale-free}} & Sparse  &  40  &0.41  & 0.41  & \textbf{0.47}  & \textbf{0.47}  & 0.41   \\ 
                    &                         & Sparse  &  700  &0.87  & 0.87  & \textbf{0.89}  & 0.82  & 0.77   \\ 
\cmidrule{1-9} 
\multirow{10}{*}{10}&\multirow{4}{*}{\textsf{Random}}  & Sparse  &  20  &0.52  & 0.52  & \textbf{0.54}  & 0.52  & 0.52  \\ 
                    &                         & Sparse  &  350  &0.91  & 0.91  & \textbf{0.92}  & 0.89  & 0.89  \\ 
                    &                         & Dense  &  20  &0.68  & \textbf{0.69}  & \textbf{0.69}  & 0.68  & 0.68   \\ 
                    &                         & Dense  &  350  &\textbf{0.94}  & 0.93  & 0.93  & 0.93  & 0.91   \\ 
\cmidrule{2-9} 
                    &\multirow{4}{*}{\textsf{Cluster}} & Sparse  &  20  &\textbf{0.50}  & \textbf{0.50}  & \textbf{0.50}  & \textbf{0.50}  & 0.48   \\ 
                    &                         & Sparse  &  350  &0.84  & 0.85  & 0.85  & 0.84  & \textbf{0.86}   \\ 
                    &                         & Dense  &  20  &0.81  & 0.81  & \textbf{0.83}  & 0.81  & 0.77   \\ 
                    &                         & Dense  &  350  &\textbf{0.95}  & \textbf{0.95}  & \textbf{0.95}  & 0.94  & 0.94   \\ 
\cmidrule{2-9} 
                 &\multirow{2}{*}{\textsf{Scale-free}} & Sparse  &  20  &0.61  & 0.61  & \textbf{0.65}  & \textbf{0.65}  & 0.60   \\ 
                    &                         & Sparse  &  350  &0.90  & \textbf{0.91}  & \textbf{0.91}  & 0.89  & 0.87   \\ 
\bottomrule
\end{tabular}
\caption{\textit{
AUC-PR scores of the algorithms for different instances. The values are averages over 16 replications for $p \in \{10, 100\}$ and over 8 replications for $p=1000$. For each setting, the best-performing algorithm is highlighted in \textbf{bold}.}}
\label{table:AUC_PR}
\end{table}


In summary, for large-scale problems (graphs with $p = 1000$), the BD-MPL algorithm achieves significantly lower computational costs while maintaining high accuracy compared to state-of-the-art methods (BD, SS, and B-CON). BD-MPL also has several advantages over RJ-MPL, as it converges faster (Figure \ref{fig:sim_convergence_AUC-PR}) and its computationally intensive components can be parallelized (Section \ref{sec: BDMCMC}). This makes BD-MPL particularly well suited for large-scale problems with $p = 1000$ or more. For medium-scale problems (around 100 nodes), BD and BD-MPL have similar and higher accuracy than other methods (Table \ref{table:AUC_PR}), but BD-MPL is more efficient, reducing computational cost from one–two hours to under three minutes (Table \ref{table:time-AUC-PR}). For small-scale problems (around 10 nodes), BD achieves slightly higher accuracy in most cases. Overall, we recommend BD-MPL for large- and medium-scale problems and BD or SS for small-scale problems.

\section{Applications}
\label{sec:application}

We apply our proposed BD-MPL and RJ-MPL algorithms to two real-world data sets: a medium-scale data set (with 100 variables) in Section \ref{subsec:mediumdata} and a large-scale data set (with 623 variables) in Section \ref{subsec:large-scale data}. We report results from the BD-MPL algorithm, as both algorithms target the same pseudo-posterior distribution and produce virtually identical outputs when run for a sufficient number of iterations. For both datasets, our approach provides point estimates of the conditional dependence structure. Moreover, we illustrate how the method facilitates uncertainty quantification over graph structures and key graph characteristics, extending beyond traditional point estimation.

\subsection{Application to Human Gene Expression}
\label{subsec:mediumdata}

We apply the BD-MPL algorithm to infer a human gene network. The dataset consists of genetic data for $p=100$ genes from \(n=60\) unrelated individuals and is available in the \textsf{BDgraph} R package \citep{BDgraph}. Detailed information on data collection can be found in \citet{Stranger2007} and \citet{Bhadra2013}. Several other Bayesian structure learning methods have also been applied to this dataset \citep{mohammadi2015bayesianStructure, li2019graphical, mohammadi2019bdgraph, van2022g, vogels2024bayesian}.  

Genes are specific sequences of DNA that play a critical role in the functioning of organisms. Gene expression is the process through which these genes produce proteins, which then influence various biological functions. Some proteins have direct effects on the organism, such as initiating the breakdown of food. Others serve a regulatory role by \textit{activating} other genes, leading to the production of additional proteins that further activate more genes, creating a cascade of interactions. These activation relationships can be represented in a gene network, where each node corresponds to a gene, and each edge signifies an activation relationship between two genes. Mapping and understanding these gene networks are vital for elucidating disease susceptibility, ultimately contributing to advancements in treatment and public health \citep{Stranger2007}.

\begin{figure}[!ht] 
\centering
\includegraphics[width=0.55\linewidth]{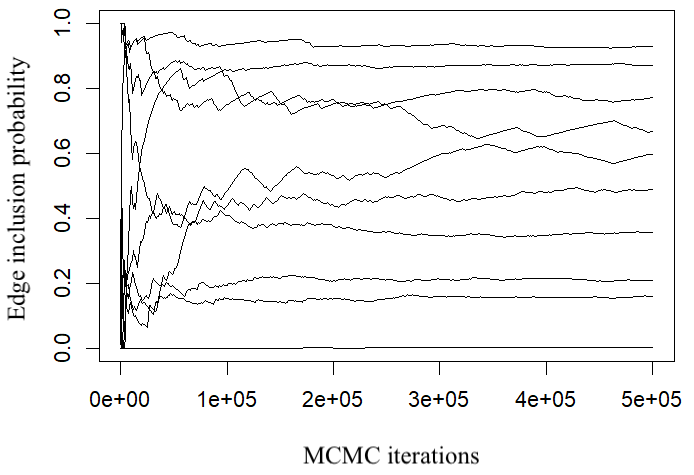}
\caption{\textit{Edge inclusion probabilities across MCMC iterations for $10$ randomly selected edges, created by the BD-MPL algorithm on the human gene dataset $(p=100)$.}}
\label{fig:trace_plot}
\end{figure}

Before applying the algorithm, we normalized the dataset using a transformation based on cumulative distribution functions \citep{Liu2009}, implemented via the \textsf{bdgraph.npn()} function in the \textsf{BDgraph} package. The initial parameter settings matched those used in the simulation study described in Section \ref{sec:simulation}. For the graph prior defined in \eqref{eq:p(g)}, we set $\beta = 0.2$, building on the results of earlier studies \citep{van2022g}. The MCMC procedure was initialized at the empty graph. To determine the number of MCMC iterations, we use a trace plot (Figure \ref{fig:trace_plot}). Trace plots are a common tool to determine MCMC convergence, see for example \citet{van2022g}. They depict the edge inclusion probabilities for a set of randomly selected edges across MCMC iterations. Based on Figure \ref{fig:trace_plot}, we observe that the BD-MPL algorithm typically converges after approximately 500,000 iterations. We therefore set the number of iterations to one million, half of which are burn-in iterations. The total computation time remained under 10 minutes on a standard desktop computer. 

\begin{figure}[!ht] 
    \centering
    \begin{tabular}[t]{cc}
        \begin{subfigure}[t]{0.5\textwidth}
            \centering
            \includegraphics[width=0.8\linewidth]{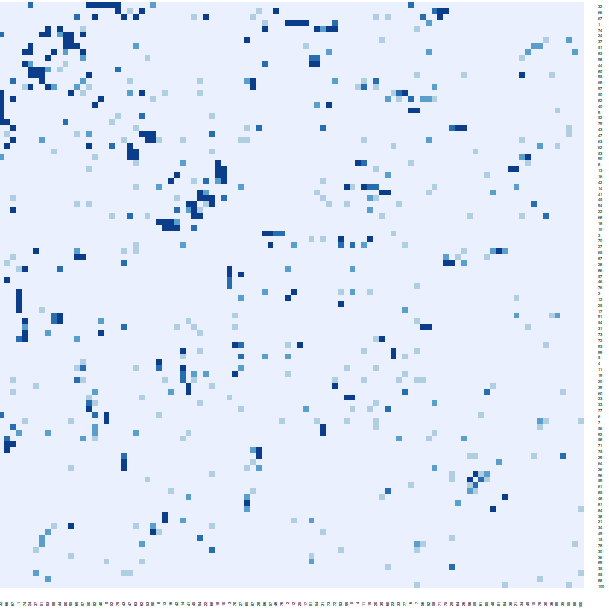}
        \end{subfigure} &
        \begin{subfigure}[t]{0.5\textwidth}
            \centering
                \includegraphics[width=0.85\textwidth]{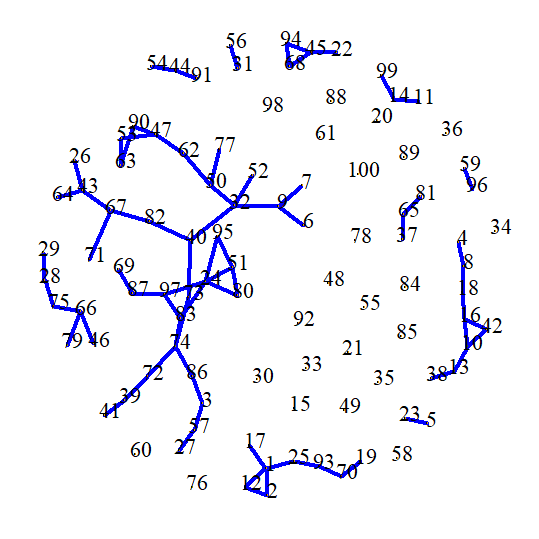} 
        \end{subfigure}
    \end{tabular}\\
\caption{\textit{(Left) Heatmap of estimated edge inclusion probabilities from the BD-MPL algorithm for the human gene dataset ($p=100$), with values ranging from $0$ (gray) to $1$ (dark blue). (Right) Estimated graph from the BD-MPL algorithm for the same dataset, showing only edges with inclusion probabilities greater than $0.9$.
}}
\label{fig:gene_networks_heatmaps}
\end{figure}

Figure \ref{fig:gene_networks_heatmaps} (left) presents a heatmap of the estimated edge inclusion probabilities \eqref{eq:p_ij} obtained from the BD-MPL algorithm. Figure \ref{fig:gene_networks_heatmaps} (right) presents a point estimate of the gene network obtained using a threshold of 0.9 on the edge inclusion probabilities. We use this relatively high threshold to prevent the plot from becoming overly dense and difficult to interpret. However, for optimal point estimation, a threshold of 0.5 is generally recommended, as suggested by \citet{barbieri2004optimal}. The inferred gene network displays structural patterns consistent with those reported in \citet[Figure 4]{Bhadra2013}. 

\begin{figure}[!ht] 
\centering
\includegraphics[width=0.65\linewidth]{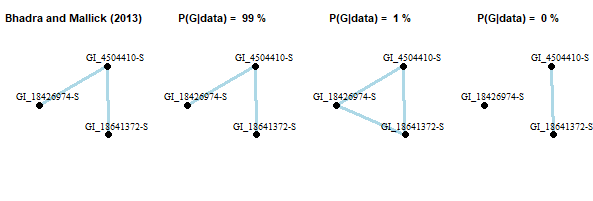}
\caption{\textit{The structure estimated by \citet[Figure 4]{Bhadra2013} among three genes part of the major histocompatibility complex (left), alongside the three most likely structures estimated by the BD-MPL.}}
\label{fig:substructure_3genes}
\end{figure}

\begin{figure}[!ht] 
\centering
\includegraphics[width=0.5\linewidth]{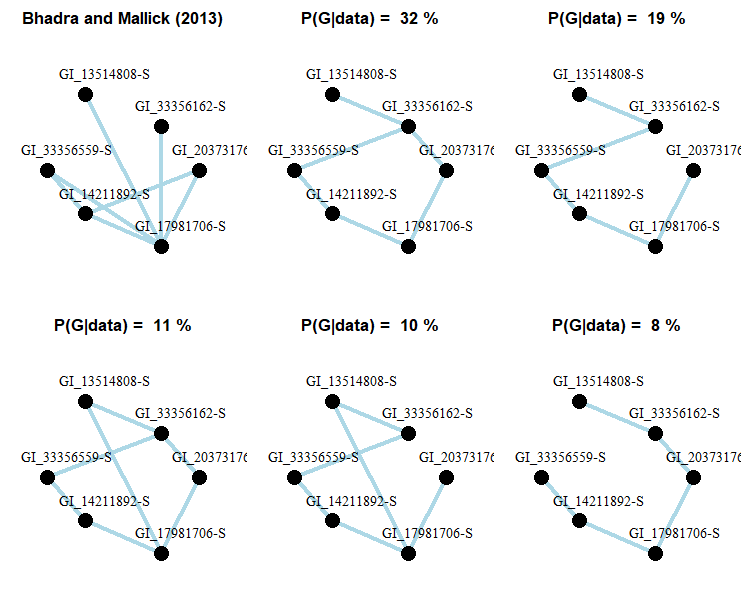}
\caption{\textit{The structure estimated by \citet[Figure 4]{Bhadra2013} for six selected genes (top left), alongside the five most likely structures estimated by the BD-MPL.}}
\label{fig:substructure_6genes}
\end{figure}

\begin{figure}[!ht] 
\centering
\includegraphics[width=0.45\linewidth]{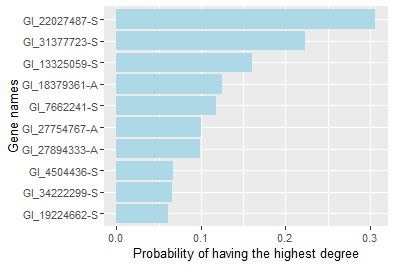}
\caption{\textit{The ten most connected genes in the human gene dataset ($p=100$), along with their corresponding posterior probabilities of being the gene with the highest degree.}}
\label{fig:gene_node_degree}
\end{figure}

In addition to estimating edge inclusion probabilities, the BD-MPL algorithm can quantify uncertainty in various graph structures. Specifically, they estimate the expected value \(E(f(G)|\Xb)\) for any function \(f(G)\) using \eqref{eq:E_f}. \citet{Bhadra2013} examined the conditional dependence structure among three genes in the major histocompatibility complex, a DNA region involved in immune system function. The BD-MPL algorithm confirms this as the most likely structure, estimating a \(99\%\) probability given the data (Figure \ref{fig:substructure_3genes}). Additionally, \citet{Bhadra2013} identified a conditional dependence structure among six genes, five of which are located on the Y chromosome. The BD-MPL algorithm estimates a near-zero probability for this structure and instead suggests alternative likely structures for these six genes (Figure \ref{fig:substructure_6genes}). Another choice of \( f(G) \) allows for identifying highly connected nodes in a network, a measure known as degree centrality, which is particularly useful for analyzing gene networks \citep{Dirk2008}. \citet{Bhadra2013} found that the gene with the highest degree is GI-22027487-S, which regulates iron and copper levels \citep{Xu2022}. Our results confirm that this gene is the most likely to have the highest degree, with a posterior probability of 30\%. Figure \ref{fig:gene_node_degree} presents the highest degree probabilities for nine other nodes.  

\subsection{Application to Gene Expression in Immune Cells}
\label{subsec:large-scale data}

This subsection presents an application of the BD-MPL algorithm on a large dataset with \(p=623\) variables and \(n=653\) observations, demonstrating its ability to estimate uncertainty across a wide range of graph structures. The dataset is the GSE15907 microarray dataset from \citet{Painter2011} and \citet{Desch2011}, comprising gene expression data from 24,922 genes in 653 mouse immune cells, obtained from the Immunological Genome Project \citep{ImmGen}. It was previously analyzed by \citet{chandra2024bayesian} using Bayesian methods. These immune cells play a crucial role in defending organisms against diseases, with genes producing proteins that either directly impact disease response or initiate cascades by activating other genes. These activation relationships form gene networks, where nodes represent genes and edges denote activation connections. Understanding these networks is essential for advancing research on immune system function and disease treatment strategies.  

\begin{figure}[!ht] 
\centering
     \includegraphics[width=0.9\linewidth]{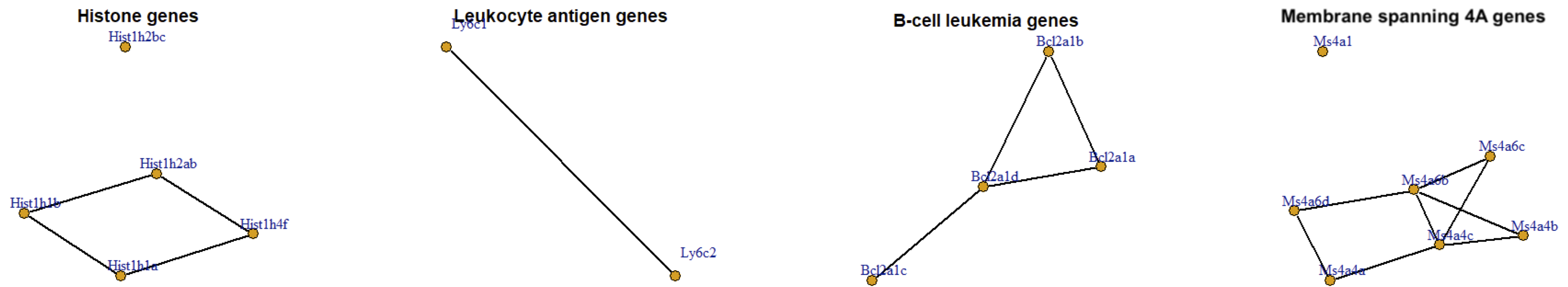}
    \caption{\textit{The networks for histone genes, leukocyte antigen genes, B-cell leukemia genes, and membrane-spanning 4A genes, estimated by the BD-MPL for the mice gene dataset ($p=623$). All displayed edges have inclusion probabilities exceeding 99\%.}}
    \label{fig:gene_connections_mice}
\end{figure}

For data preparation, following \citet{chandra2024bayesian}, we apply a $\log_2$ transformation and retain the top 2.5\% of genes with the highest variance. To normalize the dataset, we use a cumulative distribution function-based transformation, as described in Subsection \ref{subsec:mediumdata}. For the graph prior defined in \eqref{eq:p(g)}, we select $\beta = 0.01$, based on the results of \citet{chandra2024bayesian}. To determine the number of MCMC iterations, creating a trace-plot as in Figure \ref{fig:trace_plot} requires excessive memory for large-scale instances. Instead, we look at results from our simulation study in Section \ref{sec:simulation} (see Table 7 in the Supplementary Material). For the case $p = 1000$, these results suggest that the number of MCMC iterations depends on the sparsity and the number of observations, with dense and low-observation instances requiring more MCMC iterations. In all cases though, AUC-PR convergence happens within 1.5 million MCMC iterations. To be on the safe side we run 4 million MCMC iterations, discarding the first half as burn-in. The total runtime is approximately 17 hours. All other initial parameters are consistent with those used in the simulation study. We provide a heatmap of edge inclusion probabilities estimated by the BD-MPL algorithm in the supplementary material. While thresholding these probabilities could produce inferred gene networks, we do not present the complete networks here because of their large size and limited interpretability.
Instead, following \citet{chandra2024bayesian}, we focus on specific gene subsets documented in the literature as conditionally dependent. These include histone genes (e.g., Hist1h1a, Hist1h1b) \citep{WOLFFE2001948}, B-cell leukemia genes (e.g., Bcl2a1a, Bcl2a1b) \citep{chandra2024bayesian}, leukocyte antigen genes (e.g., Ly6c1, Ly6c2) \citep{lee2013ly6}, and membrane-spanning 4A genes (e.g., Ms4a1, Ms4a4c) \citep{liang2001structural}. Figure \ref{fig:gene_connections_mice} highlights these subsets, showing edges with inclusion probabilities greater than 0.99. These results demonstrate that the BD-MPL algorithm successfully recovers known conditional dependencies within biologically relevant gene groups.

\begin{figure}[!ht] 
\centering
     \includegraphics[width=0.8\linewidth]{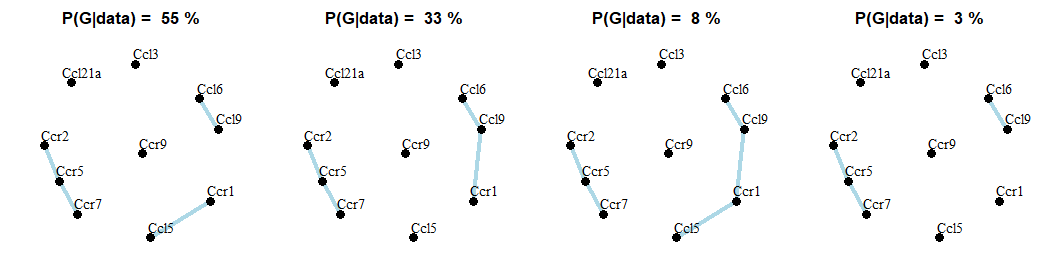}
    \caption{\textit{The most likely structures between chemokine ligand (Ccl) and chemokine receptor (Ccr) genes, estimated by the BD-MPL on the mice gene dataset ($p=623$).}}
    \label{fig:cc_chemokines}
\end{figure}

\begin{figure}[!ht] 
\centering
     \includegraphics[width=0.55\linewidth]{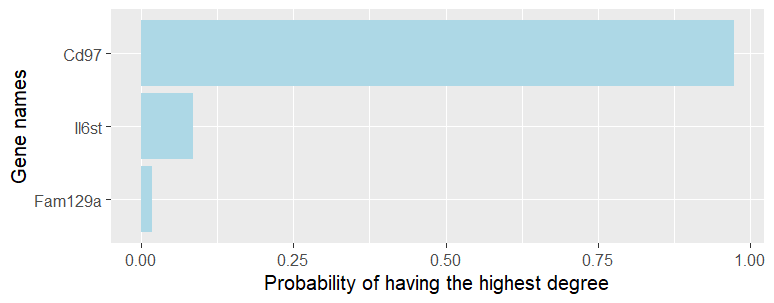}
    \caption{\textit{The three most connected genes in the mice gene dataset ($p=623$), with their corresponding posterior probability of being the gene with the highest degree.}}
    \label{fig:mice_gene_degree}
\end{figure}

To illustrate the ability of BD-MPL to quantify uncertainty in specific graph structures, we focus on a group of genes encoding chemokine ligands (e.g., Ccl3, Ccl5) and chemokine receptors (e.g., Ccr2, Ccr5). These genes produce CC chemokines, which play a crucial role in immune responses \citep{Raman2011}. Figure \ref{fig:cc_chemokines} presents the four most likely conditional dependence structures among these genes. We also find that the CD97 gene has the highest degree with a posterior probability of 97\% (Figure \ref{fig:mice_gene_degree}). This aligns with the conclusion of \citet{Safaee2013} that the CD97 protein, encoded by this gene, is broadly expressed and plays diverse roles in the immune system.  

\section{Conclusion}
\label{sec:concl}

We address the computational challenges of posterior sampling for GGMs in large-scale problems by introducing an MCMC-based framework that combines birth–death and reversible jump algorithms with the marginal pseudo-likelihood approach. The proposed algorithms scale to large graph spaces, enabling parallel exploration for graphs with over 1,000 nodes, as demonstrated in our extensive simulation study. We provide theoretical guarantees, including posterior contraction and graph selection consistency. In addition to scalability and theoretical support, the proposed method offers reliable uncertainty quantification and the flexibility to incorporate prior knowledge about the graph structure, as illustrated in our application section.




Future research could enhance these methods along several directions. In Section \ref{sec:MPL}, we employ a fractional prior to derive a closed-form analytical expression for the local components, see Equation \eqref{eq:mpl_3}. One potential enhancement is to tune the prior to improve inferential accuracy and improve the approximation of the true posterior. Alternatively, one could adopt the closed-form approximation for the ratio of normalizing constants proposed by \citet{mohammadi2023accelerating}. Another promising avenue is the exploration of alternative priors on the precision matrix. For instance, priors such as the spike-and-slab formulation used in \citet{jalali2020b} may offer improved structure learning and posterior inference.

The BD-MPL and RJ-MPL algorithms rely on the addition or removal of a single edge per MCMC iteration to explore the graph space. Their computational efficiency can be further improved by allowing multiple edge updates per iteration. For instance, for the BD-MPL algorithm, after computing all possible edge update probabilities from \eqref{eq:prob bd}, one can select a fixed number of edges ($k > 1$) to update in each iteration. The choice of $k$ can be guided by the computational time required per iteration. This strategy is implemented in the \textsf{BDgraph} package \citep{BDgraph} through the function \textsf{bdgraph.mpl()} using the option \textsf{jump = k}; see \citet[Section 5.5]{dobra2015} for further details.

Another promising direction is to adopt a blocking approach, as proposed by \cite{vandenboom2023} and \cite{colombi2024}. In particular, the block-Bernoulli prior introduced in \cite{vandenboom2023} enables joint updates of groups of edges by operating at the block level. Analogous to \eqref{eq:mpl_3}, local pseudo-likelihoods for blocks can be expressed in closed form using Equation (29) from \cite{consonni2012objective}. Since the proposed algorithm assumes that each block is either fully connected or empty, the corresponding block marginal likelihoods can be computed exactly.

Scalability is a major barrier to applying Bayesian inference in graphical models. In this work, we address this challenge by proposing a scalable Bayesian framework based on the marginal pseudo-likelihood approach for vanilla (single) GGMs. A natural extension of this framework is to the multiple GGMs setting \cite{peterson2015bayesian}. A related extension has been developed by \cite{jalali2023bayesian}, who proposed a Bayesian method for multiple GGMs using a jointly convex, regression-based pseudo-likelihood with a spike-and-slab prior. More recently, \cite{avalos2025bayesian} applied a similar pseudo-likelihood framework to scale Bayesian inference for multiple Ising graphical models with binary data. For possible extensions of our proposed algorithms to multiple Ising or general discrete graphical models, we refer readers to \citet[Section 5]{dobra2015}, which employs a birth-death MCMC algorithm with an MPL approach for multivariate discrete data.

\makeatletter
\renewcommand \thesection{S\@arabic\c@section}
\renewcommand\thetable{S\@arabic\c@table}
\renewcommand \thefigure{S\@arabic\c@figure}
\makeatother
\bigskip
\begin{center}
{\large\bf SUPPLEMENTARY MATERIAL}
\end{center}

\begin{description}

\item[R-package:] 
The R package \textsf{BDgraph} contains code implementing our method described in this article. The BD-MPL and RJ-MPL algorithms are implemented in the function \textsf{bdgraph.mpl()}. The package is freely available from the Comprehensive R Archive Network (CRAN) at \url{http://cran.r-project.org/packages=BDgraph}. 

\item[GitHub repository:] 
The code for reproducing all the results from our simulation study in Section \ref{sec:simulation} and our applications in Section \ref{sec:application}, as well as instructions on how to download and process the data for analyses, is available on the GitHub page at \url{https://github.com/lucasvogels33/Large-scale-BSL-for-GGMs-using-MPL}. 

\item[Data sets:] 
The data set used in Subsection \ref{subsec:mediumdata} is available in the R package \textsf{BDgraph} (geneExpression.RData file). The data set used in Subsection \ref{subsec:large-scale data} can be found on the GitHub page linked above (cleaned\_data.Rdata file). Additionally, detailed information on data processing is provided on the GitHub page. 

\item[Supplementary materials:]
The supplementary materials provide additional results for the simulations presented in Section \ref{sec:simulation} and the applications in Section \ref{sec:application}.  Section \ref{sec:simulation} presents supplementary results for the simulations discussed in the manuscript. Section \ref{subsec:mediumdata} contains further results for the real-world application to human Gene expression covered in the manuscript. Section \ref{subsec:large-scale data} provides additional results for the real-world application to Gene expression in immune cells described in the manuscript.

\end{description}

\section{Additional Materials for Simulation Study}
\label{sec:simulation}

\begin{figure}[t]
\centering
\includegraphics[width=1\linewidth]{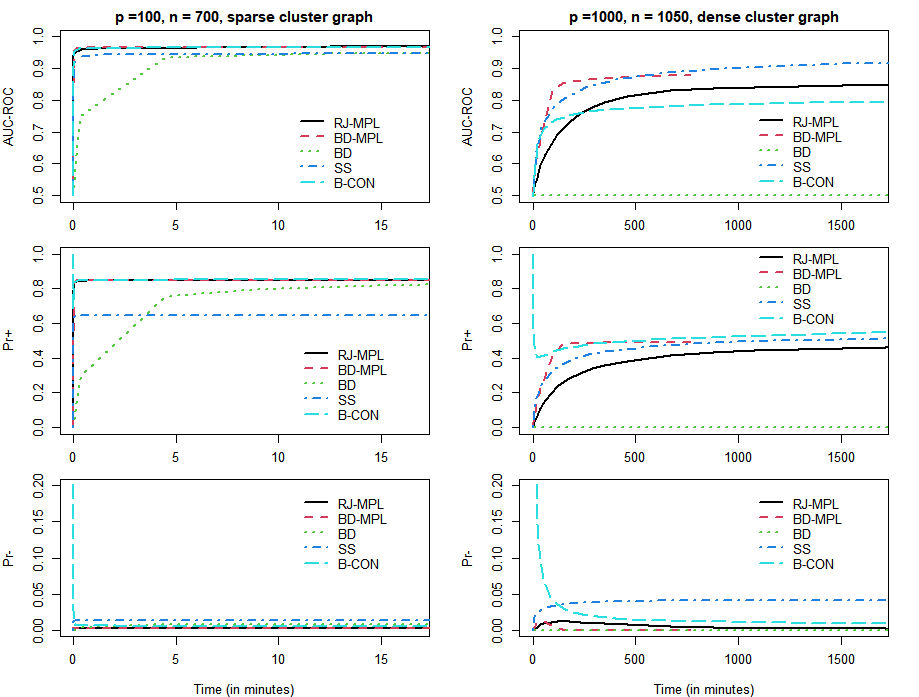}
\caption{\textit{
The convergence of AUC-ROC (top row), $Pr+$ (middle row), and $Pr-$ (bottom row) scores over running time for all algorithms (RJ-MPL, BD-MPL, BD, SS, B-CON). The plots on the left represent the instance with \(p = 100\), \(n = 700\) for the sparse \textsf{\textsf{Cluster}} graph. The plots on the right represent the instance with \(p = 1000\), \(n = 1050\) for the dense \textsf{\textsf{Cluster}} graph.}}
\label{fig:convergence_AUC-ROC_Pr}
\end{figure}

Here we present additional simulation results: the edge density of the simulated graphs in Table \ref{table:edges}, four additional graph precision recovery metrics, and the graph precision recovery metrics over time in Figure \ref{fig:convergence_AUC-ROC_Pr}.

The Area Under the Receiver Operating Characteristic Curve (AUC-ROC) \citep{hanley1982} evaluates a classifier's ability to distinguish between true edges and non-edges in the graph. The ROC curve plots the True Positive Rate against the False Positive Rate at various threshold settings. The AUC-ROC measures the area under the curve, ranging from 0 to 1, with higher values indicating better performance. Table \ref{table:AUC} presents the AUC-ROC scores, and Table \ref{table:time-AUC-ROC} reports the computational time required for AUC-ROC convergence. Here, AUC-ROC convergence is defined as the time at which the AUC-ROC value reaches a value within 0.01 of its final iteration value. 

The $F1$ Score \citep{powers2020evaluation} is the harmonic mean of Precision and Recall, providing a single metric that balances both. It is defined as:
\begin{equation}
\label{eq:f1}
F1 = 2 \times \frac{{\text{Precision} \times \text{Recall}}}{{\text{Precision} + \text{Recall}}}.
\end{equation}
To report the $F1$ values, we first obtain the estimated graph \(\hat{G} = (V, \hat{E})\), where \(\hat{E} = \left\{e = (i, j) \mid \hat{P}_e \geq 0.5\right\}\). In Bayesian graphical learning, the $F1$ Score ranges from 0 to 1, with higher values indicating better overall performance in detecting true edges while minimizing false positives and false negatives. Table \ref{table:F1} presents the $F1$ scores.

$Pr^+$ and $Pr^-$ represent the average inclusion probability for all edges and non-edges, respectively, in the true graph $G = (V, E)$ \citep{vogels2024bayesian}. They are calculated as:
\begin{equation}
\label{eq:pr+}
Pr^+ =\frac{1}{|E|} \sum_{e \in E} \hat{P}_e
\end{equation}
and
\begin{equation}
\label{eq:pr-}
Pr^- =\frac{1}{|\bar{E}|} \sum_{e \in \bar{E}} \hat{P}_e,
\end{equation}
where $\hat{P}_e$ are the estimated edge-inclusion probabilities of the manuscript. These probabilities serve as measures of calibration accuracy. Ideally, algorithms should achieve a high $Pr^+$ to enhance edge detection accuracy and a low $Pr^-$ to effectively reject edges not present in the true graph $G = (V, E)$. We report the $Pr^+$ values in Table \ref{table:pplus} and $Pr^-$ in Table \ref{table:pmin}.

In summary, for AUC-ROC and \(F1\) metrics, the RJ-MPL and BD-MPL methods perform as well as or better than other algorithms. For \(Pr^+\) at \(p=100\) and \(p=1000\) (Table \ref{table:pplus}), B-CON occasionally shows higher values, likely due to its higher \(Pr^-\) values. Generally, our MPL approaches (RJ-MPL and BD-MPL) perform well in terms of AUC-PR, F1, and \(Pr^-\), but not as well for \(Pr^+\). This tendency is likely because our methods tend to select sparser graphs compared to other approaches. Ideally, we aim for a high \(Pr^+\) to improve edge detection accuracy while maintaining a low \(Pr^-\) to effectively reject non-edges.


\begin{table}[t]
\centering
\begin{tabular}{l l l l l l l l l l}
\arrayrulecolor{blue}\toprule
$p$                 &&\multicolumn{2}{c}{10}    && \multicolumn{2}{c}{100} && \multicolumn{2}{c}{1000}\\
\cmidrule{3-4}
\cmidrule{6-7}
\cmidrule{9-10}
Density             && Sparse & Dense           && Sparse & Dense          && Sparse & Dense\\
\hline 
\textsf{\textsf{Random}}     && 11.1\% & 44.4\%          && 1.0\% & 5.0\%           && 0.5\% & 5.0\%\\
\textsf{\textsf{Cluster}}    && 11.1\% & 44.4\%          && 1.0\% & 5.0\%           && 0.5\% & 5.0\%\\
\textsf{\textsf{Scale-free}} &&\multicolumn{2}{c}{20.0\%}&&\multicolumn{2}{c}{2.0\%}&& \multicolumn{2}{c}{0.2\%}\\
\bottomrule
\end{tabular}
\caption{\textit{The edge density of the graphs is defined as the proportion of the number of edges to the total number of possible edges in the graphs.}}
\label{table:edges}
\end{table}

\vspace*{\fill}
\begin{table}[ht] 
\centering
\begin{tabular}{c @{\hspace{0.7em}} c  c c @{\hspace{2em}} c c c @{\hspace{1.7em}} c @{\hspace{0.7em}} c}
\arrayrulecolor{blue}\toprule
$p$   & Graph  & Density & $n$ &  RJ-MPL & BD-MPL & BD & SS & B-CON  \\ 
\hline 
\multirow{10}{*}{1000}&\multirow{4}{*}{\textsf{Random}}& Sparse &  400  & 1287  & \textbf{45}  & - &321  & 664   \\ 
                    &                         & Sparse &  1050 & 851  & \textbf{79}  & - & 328  & 338   \\ 
                    &                         & Dense  &  400  & 2581  & \textbf{499}  & - & 2097  & 3528   \\ 
                    &                         & Dense  &  1050 & 2125  & \textbf{901}  & - & 2122  & 1752   \\ 
\cmidrule{2-9} 
                    &\multirow{4}{*}{\textsf{Cluster}} & Sparse &  400  & 1374  & \textbf{38}  & - & 301  & 633  \\ 
                    &                         & Sparse &  1050 & 874  & \textbf{71}  & - &318  & 240   \\ 
                    &                         & Dense  &  400  & 2429  & \textbf{480}  & - &2111  & 1626   \\ 
                    &                         & Dense  &  1050 & 2160  & \textbf{402}  & - &1791  & 916   \\ 
\cmidrule{2-9} 
                  &\multirow{2}{*}{\textsf{Scale-free}}& Sparse &  400  & 2143  & \textbf{19}  & - &387  & 595  \\ 
                    &                         & Sparse &  1050 & 1173  & \textbf{38}  & - &1086  & 67   \\ 
\cmidrule{1-9} 
\multirow{10}{*}{100}&\multirow{4}{*}{\textsf{Random}} & Sparse  &  40   &1  & 2  & 63  & \textbf{0}  & 3  \\ 
                     &                        & Sparse  &  700  &1  & \textbf{0}  & 31  & \textbf{0}  & \textbf{0}  \\ 
                     &                        & Dense   &  40   &4  & 1  & 89  & 1  & 4   \\ 
                     &                        & Dense   &  700  &3  & \textbf{0}  & 50  & 1  & 1   \\ 
\cmidrule{2-9} 
                     &\multirow{4}{*}{\textsf{Cluster}}& Sparse  &  40   &2  & 1  & 60  & \textbf{0}  & 3   \\ 
                     &                        & Sparse  &  700  &1  & 1  & 49  & 5  & \textbf{0}   \\ 
                     &                        & Dense   &  40   &4  & 1  & 107  & \textbf{0}  & 3   \\ 
                     &                        & Dense   &  700  &2  & \textbf{1}  & 54  & \textbf{1}  & \textbf{1}   \\ 
\cmidrule{2-9} 
                 &\multirow{2}{*}{\textsf{Scale-free}} & Sparse  &  40   &5  & 4  & 85  & \textbf{0}  & 3   \\ 
                     &                        & Sparse  &  700  &3  & \textbf{0}  & 54  & 3  & \textbf{0}   \\ 
\bottomrule
\end{tabular}
\caption{
\textit{Computational cost (\(T\)) in minutes until AUC-ROC convergence for various instances. \(T\) represents the average time until AUC-ROC convergence, based on 16 replications for \(p \in \{10, 100\}\) and 8 replications for \(p = 1000\). The table excludes the \(p=10\) case since the computational time for all algorithms was less than one minute. A “-” indicates that an algorithm did not converge within five days. For each setting, the best-performing algorithm is highlighted in \textbf{bold}.}}
\label{table:time-AUC-ROC}
\end{table}
\vspace*{\fill}

\vspace*{\fill}
\begin{table}[ht]
\centering
\begin{tabular}{c @{\hspace{0.7em}} c @{\hspace{0.7em}} c c @{\hspace{1.6em}} c c c @{\hspace{1.7em}} c @{\hspace{0.7em}} c}
\arrayrulecolor{blue}\toprule
$p$   & Graph  & Density & $n$ &  RJ-MPL & BD-MPL & BD & SS & B-CON  \\ 
\hline 
\multirow{10}{*}{1000}&\multirow{4}{*}{\textsf{Random}}& Sparse  &  400  &0.87  & 0.89  & 0.50  & \textbf{0.90}  & 0.89   \\ 
                    &                         & Sparse  &  1050  &0.91  & 0.92  & 0.50  & 0.92  & \textbf{0.94}   \\ 
                    &                         & Dense  &  400  &0.70  & 0.74  & 0.50  & \textbf{0.76}  & 0.70   \\ 
                    &                         & Dense  &  1050  &0.77  & 0.80  & 0.50  & \textbf{0.83}  & 0.78   \\ 
\cmidrule{2-9} 
                     &\multirow{4}{*}{\textsf{Cluster}}& Sparse  &  400  &0.88  & \textbf{0.90}  & 0.50  & \textbf{0.90}  & \textbf{0.90}   \\ 
                    &                         & Sparse  &  1050  &0.92  & 0.93  & 0.50  & 0.92  & \textbf{0.94}   \\ 
                    &                         & Dense  &  400  &0.78  & 0.84  & 0.50  & \textbf{0.89}  & 0.72   \\ 
                    &                         & Dense  &  1050  &0.86  & 0.88  & 0.50  & \textbf{0.93}  & 0.80   \\ 
\cmidrule{2-9} 
                  &\multirow{2}{*}{\textsf{Scale-free}}& Sparse  &  400  &0.89  & 0.90  & 0.50  & \textbf{0.92}  & 0.91   \\ 
                    &                         & Sparse  &  1050  &0.93  & 0.93  & 0.50  & 0.93  & \textbf{0.95}   \\ 
\cmidrule{1-9} 
\multirow{10}{*}{100}&\multirow{4}{*}{\textsf{Random}} & Sparse  &  40  &0.85  & 0.86  & 0.86  & \textbf{0.87}  & 0.86   \\ 
                    &                         & Sparse  &  700  &\textbf{0.97}  & \textbf{0.97}  & \textbf{0.97}  & 0.96  & \textbf{0.97}   \\ 
                    &                         & Dense  &  40  &0.75  & 0.75  & 0.75  & \textbf{0.77}  & 0.76   \\ 
                    &                         & Dense  &  700  &\textbf{0.94}  & \textbf{0.94}  & \textbf{0.94}  & 0.92  & \textbf{0.94}   \\ 
\cmidrule{2-9} 
                     &\multirow{4}{*}{\textsf{Cluster}}& Sparse  &  40  &\textbf{0.85}  & \textbf{0.85}  & 0.84  & \textbf{0.85}  & \textbf{0.85}   \\ 
                    &                         & Sparse  &  700  &\textbf{0.97}  & \textbf{0.97} & 0.96  & 0.95  & \textbf{0.97}   \\ 
                    &                         & Dense  &  40  &0.77  & 0.77  & 0.77  & \textbf{0.79}  & 0.77   \\ 
                    &                         & Dense  &  700  &0.94  & \textbf{0.95}  & \textbf{0.95}  & 0.92  & 0.94   \\ 
\cmidrule{2-9} 
                  &\multirow{2}{*}{\textsf{Scale-free}}& Sparse  &  40  &0.81  & 0.80  & 0.82  & \textbf{0.84}  & 0.81   \\ 
                    &                         & Sparse  &  700  &0.95  & 0.95  & \textbf{0.96}  & 0.95  & 0.95   \\ 
\cmidrule{1-9} 
\multirow{10}{*}{10}  &\multirow{4}{*}{\textsf{Random}}& Sparse  &  20  &\textbf{0.80}  & \textbf{0.80}  & \textbf{0.80}  & 0.78  & 0.75   \\ 
                    &                         & Sparse  &  350  &0.95  & \textbf{0.96}  & \textbf{0.96}  & 0.92  & 0.94   \\ 
                    &                         & Dense  &  20  &0.68  & 0.68  & \textbf{0.69}  & 0.68  & 0.68   \\ 
                    &                         & Dense  &  350  &\textbf{0.92}  & \textbf{0.92}  & \textbf{0.92}  & 0.91  & 0.90   \\ 
\cmidrule{2-9} 
                     &\multirow{4}{*}{\textsf{Cluster}}& Sparse  &  20  &0.73  & \textbf{0.75}  & \textbf{0.75}  & \textbf{0.75}  & 0.74   \\ 
                    &                         & Sparse  &  350  &0.90  & 0.91  & 0.91  & 0.90  & \textbf{0.92}   \\ 
                    &                         & Dense  &  20  &0.81  & 0.81  & \textbf{0.82}  & 0.79  & 0.75   \\ 
                    &                         & Dense  &  350  &0.94  & 0.94  & \textbf{0.95}  & 0.92  & 0.92   \\ 
\cmidrule{2-9} 
                  &\multirow{2}{*}{\textsf{Scale-free}}& Sparse  &  20  &0.76  & 0.76  & 0.78  & \textbf{0.79}  & 0.76   \\ 
                    &                         & Sparse  &  350  &0.92  & \textbf{0.94}  & \textbf{0.94}  & 0.92  & 0.92   \\ 
\bottomrule
\end{tabular}
\caption{\textit{
$AUC-ROC$ scores of the algorithms for different instances. The $AUC-PR$ reaches its best score at $1$ and its worst at $0$. The values are averages over 16 replications for $p \in \{10, 100\}$ and over 8 replications for $p=1000$. For each setting, the best-performing algorithm is highlighted in \textbf{bold}.}}
\label{table:AUC}
\end{table}
\vspace*{\fill}

\vspace*{\fill}
\begin{table}[ht]
\centering
\begin{tabular}{c @{\hspace{0.7em}} c @{\hspace{0.7em}} c c @{\hspace{1.6em}} c c c @{\hspace{1.7em}} c @{\hspace{0.7em}} c}
\arrayrulecolor{blue}\toprule
$p$   & Graph  & Density & $n$ &  RJ-MPL & BD-MPL & BD & SS & B-CON  \\ 
\hline 
\multirow{10}{*}{1000}&\multirow{4}{*}{\textsf{Random}}& Sparse  &  400  &\textbf{0.73}  & \textbf{0.73}  & 0.00  & 0.68  & 0.65   \\ 
                    &                         & Sparse  &  1050  &\textbf{0.84}  & \textbf{0.84}  & 0.00  & 0.75  & 0.72   \\ 
                    &                         & Dense  &  400  &\textbf{0.40}  & \textbf{0.40}  & 0.00  & 0.39  & \textbf{0.40}   \\ 
                    &                         & Dense  &  1050  &0.59  & \textbf{0.60}  & 0.00  & 0.55  & 0.56   \\ 
\cmidrule{2-9} 
                     &\multirow{4}{*}{\textsf{Cluster}}& Sparse  &  400  &\textbf{0.75}  & \textbf{0.75}  & 0.00  & 0.69  & 0.70   \\ 
                    &                         & Sparse  &  1050  &\textbf{0.85}  & \textbf{0.85}  & 0.00  & 0.76  & 0.75   \\ 
                    &                         & Dense  &  400  &0.47  & 0.47  & 0.00  & 0.47  & \textbf{0.55}   \\ 
                    &                         & Dense  &  1050  &0.65  & 0.65  & 0.00  & 0.63  & \textbf{0.67}   \\ 
\cmidrule{2-9} 
                  &\multirow{2}{*}{\textsf{Scale-free}}& Sparse  &  400  &\textbf{0.62}  & \textbf{0.62}  & 0.00  & 0.63  & 0.49   \\ 
                    &                         & Sparse  &  1050  &0.75  & 0.75  & 0.00  & \textbf{0.79}  & 0.53   \\ 
\cmidrule{1-9} 
\multirow{10}{*}{100}&\multirow{4}{*}{\textsf{Random}} & Sparse  &  40  &0.41  & 0.41  & 0.54  & \textbf{0.57}  & 0.52   \\ 
                    &                         & Sparse  &  700  &0.85  & 0.84  & \textbf{0.89}  & 0.75  & 0.79   \\ 
                    &                         & Dense  &  40  &0.38  & 0.38  & \textbf{0.42}  & 0.37  & 0.39   \\ 
                    &                         & Dense  &  700  &0.85  & 0.85  & \textbf{0.86}  & 0.65  & 0.79   \\ 
\cmidrule{2-9} 
                     &\multirow{4}{*}{\textsf{Cluster}}& Sparse  &  40  &0.44  & 0.44  & 0.52  & \textbf{0.54}  & 0.51   \\ 
                    &                         & Sparse  &  700  &0.83  & 0.83  & \textbf{0.87}  & 0.75  & 0.78   \\ 
                    &                         & Dense  &  40  &0.40  & 0.39  & \textbf{0.42}  & 0.39  & 0.41   \\ 
                    &                         & Dense  &  700  &0.85  & 0.85  & \textbf{0.86}  & 0.69  & 0.81   \\ 
\cmidrule{2-9} 
                  &\multirow{2}{*}{\textsf{Scale-free}}& Sparse  &  40  &0.41  & 0.41  & \textbf{0.50}  & 0.48  & 0.46   \\ 
                    &                         & Sparse  &  700  &0.86  & 0.86  & \textbf{0.89}  & 0.70  & 0.73   \\ 
\cmidrule{1-9} 
\multirow{10}{*}{10}  &\multirow{4}{*}{\textsf{Random}}& Sparse  &  20  &\textbf{0.40}  & \textbf{0.40}  & 0.36  & 0.27  & 0.33   \\ 
                    &                         & Sparse  &  350  &\textbf{0.9}  & \textbf{0.9}  & 0.89  & 0.55  & \textbf{0.90}   \\ 
                    &                         & Dense  &  20  &\textbf{0.37}  & \textbf{0.37}  & 0.33  & 0.24  & 0.33   \\ 
                    &                         & Dense  &  350  &0.84  & 0.84  & 0.83  & 0.6  & 0.84   \\ 
\cmidrule{2-9} 
                     &\multirow{4}{*}{\textsf{Cluster}}& Sparse  &  20  &\textbf{0.43}  & \textbf{0.43}  & 0.35  & 0.19  & 0.38   \\ 
                    &                         & Sparse  &  350  &0.83  & \textbf{0.84}  & 0.82  & 0.60  & 0.82   \\ 
                    &                         & Dense  &  20  &0.44  & 0.44  & \textbf{0.41}  & 0.28  & 0.37   \\ 
                    &                         & Dense  &  350  &0.81  & 0.81  & 0.81  & 0.69  & \textbf{0.84}   \\ 
\cmidrule{2-9} 
                  &\multirow{2}{*}{\textsf{Scale-free}}& Sparse  &  20  &0.47  & 0.47  & \textbf{0.49}  & 0.33  & 0.41   \\ 
                    &                         & Sparse  &  350  &\textbf{0.88}  & \textbf{0.88}  & 0.88  & 0.66  & 0.83   \\ 
\bottomrule
\end{tabular}
\caption{\textit{$F1$ scores (at a threshold of $0.5$) of the algorithms for different instances. The $F1$ score reaches its best score at $1$ and its worst at $0$. The values are averages over 16 replications for $p \in {10, 100}$ and over 8 replications for $p=1000$. For each setting, the best-performing algorithm is highlighted in \textbf{bold}. }}
\label{table:F1}
\end{table}
\vspace*{\fill}

\vspace*{\fill}
\begin{table}[ht]
\centering
\begin{tabular}{c @{\hspace{0.7em}} c @{\hspace{0.7em}} c c @{\hspace{1.6em}} c c c @{\hspace{1.7em}} c @{\hspace{0.7em}} c}
\arrayrulecolor{blue}\toprule
$p$   & Graph  & Density & $n$ &  RJ-MPL & BD-MPL & BD & SS & B-CON  \\ 
\hline 
\multirow{10}{*}{1000}&\multirow{4}{*}{\textsf{Random}}& Sparse  &  400  &0.63  & 0.65  & 0.00  & 0.62  & \textbf{0.67}   \\ 
                    &                         & Sparse  &  1050  &0.74  & 0.76  & 0.00  & 0.64  & \textbf{0.78}   \\ 
                    &                         & Dense  &  400  &0.26  & 0.26  & 0.00  & 0.31  & \textbf{0.34}   \\ 
                    &                         & Dense  &  1050  &0.41  & 0.44  & 0.00  & 0.44  & \textbf{0.51}   \\ 
\cmidrule{2-9} 
                     &\multirow{4}{*}{\textsf{Cluster}}& Sparse  &  400  &0.64  & 0.66  & 0.00  & 0.64  & \textbf{0.69}   \\ 
                    &                         & Sparse  &  1050  &0.75  & 0.78  & 0.00  & 0.66  & \textbf{0.8}   \\ 
                    &                         & Dense  &  400  &0.31  & 0.32  & 0.00  & 0.38  & \textbf{0.42}   \\ 
                    &                         & Dense  &  1050  &0.48  & 0.49  & 0.00  & 0.53  & \textbf{0.56}   \\ 
\cmidrule{2-9} 
                  &\multirow{2}{*}{\textsf{Scale-free}}& Sparse  &  400  &\textbf{0.70}  & \textbf{0.70}  & 0.00  & 0.68  & \textbf{0.70}   \\ 
                    &                         & Sparse  &  1050  &0.80  & \textbf{0.81}  & 0.00  & 0.68  & \textbf{0.81}   \\ 
\cmidrule{1-9} 
\multirow{10}{*}{100}&\multirow{4}{*}{\textsf{Random}} & Sparse  &  40  &\textbf{0.54}  & \textbf{0.54}  & 0.53  & 0.50  & 0.52   \\ 
                    &                         & Sparse  &  700  &\textbf{0.87}  & \textbf{0.87}  & 0.86  & 0.65  & \textbf{0.87}   \\ 
                    &                         & Dense  &  40  &0.30  & 0.30  & \textbf{0.33}  & 0.28  & 0.29   \\ 
                    &                         & Dense  &  700  &0.75  & 0.75  & 0.77  & 0.52  & \textbf{0.78}   \\ 
\cmidrule{2-9} 
                     &\multirow{4}{*}{\textsf{Cluster}}& Sparse  &  40  &\textbf{0.54}  & \textbf{0.54}  & 0.52  & 0.49  & 0.52   \\ 
                    &                         & Sparse  &  700  &\textbf{0.85}  & \textbf{0.85}  & 0.84  & 0.64  & \textbf{0.85}   \\ 
                    &                         & Dense  &  40  &0.31  & 0.31  & \textbf{0.34}  & 0.30  & 0.30   \\ 
                    &                         & Dense  &  700  &0.75  & 0.75  & \textbf{0.78}  & 0.55  & \textbf{0.78}   \\ 
\cmidrule{2-9} 
                  &\multirow{2}{*}{\textsf{Scale-free}}& Sparse  &  40  &0.40  & 0.41  & \textbf{0.43}  & 0.38  & 0.38   \\ 
                    &                         & Sparse  &  700  &0.82  & 0.82  & \textbf{0.83}  & 0.56  & 0.81   \\ 
\cmidrule{1-9} 
\multirow{10}{*}{10}  &\multirow{4}{*}{\textsf{Random}}& Sparse  &  20  &\textbf{0.36}  & \textbf{0.36}  & 0.31  & 0.24  & 0.25   \\ 
                    &                         & Sparse  &  350  &\textbf{0.84}  & \textbf{0.84}  & 0.82  & 0.44  & 0.83   \\ 
                    &                         & Dense  &  20  &\textbf{0.28}  & \textbf{0.28}  & 0.26  & 0.19  & 0.21   \\ 
                    &                         & Dense  &  350  &0.73  & 0.73  & 0.73  & 0.49  & \textbf{0.75}   \\ 
\cmidrule{2-9} 
                     &\multirow{4}{*}{\textsf{Cluster}}& Sparse  &  20  &\textbf{0.36}  & \textbf{0.36}  & 0.30  & 0.21  & 0.26   \\ 
                    &                         & Sparse  &  350  &0.74  & \textbf{0.75}  & 0.73  & 0.46  & 0.74   \\ 
                    &                         & Dense  &  20  &\textbf{0.32}  & \textbf{0.32}  & 0.3  & 0.23  & 0.25   \\ 
                    &                         & Dense  &  350  &0.69  & 0.69  & 0.69  & 0.55  & \textbf{0.73}   \\ 
\cmidrule{2-9} 
                  &\multirow{2}{*}{\textsf{Scale-free}}& Sparse  &  20  &\textbf{0.37}  & \textbf{0.37}  & 0.36  & 0.26  & 0.29   \\ 
                    &                         & Sparse  &  350  &\textbf{0.79}  & \textbf{0.79}  & \textbf{0.79}  & 0.53  & 0.78   \\ 
\bottomrule
\end{tabular}
\caption{\textit{$Pr^+$ scores of the algorithms for different instances. The $Pr^+$ reaches its best score at $1$ and its worst at $0$. The values are averages over 16 replications for $p \in \{10, 100\}$ and over 8 replications for $p=1000$. For each setting, the best-performing algorithm is highlighted in \textbf{bold}.}}
\label{table:pplus}
\end{table}
\vspace*{\fill}

\vspace*{\fill}
\begin{table}[ht]
\centering
\begin{tabular}{c @{\hspace{0.7em}} c @{\hspace{0.7em}} c c @{\hspace{1.6em}} c c c @{\hspace{1.7em}} c @{\hspace{0.7em}} c}
\arrayrulecolor{blue}\toprule
$p$   & Graph  & Density & $n$ &  RJ-MPL & BD-MPL & BD & SS & B-CON  \\ 
\hline 
\multirow{10}{*}{1000}&\multirow{4}{*}{\textsf{Random}}& Sparse  &  400  &\textbf{0.00}  & \textbf{0.00}  & \textbf{0.00}  & 0.03  & 0.01   \\ 
                    &                         & Sparse  &  1050  &\textbf{0.00}  & \textbf{0.00}  & \textbf{0.00}  & 0.02  & 0.01   \\ 
                    &                         & Dense  &  400  &\textbf{0.00}  & \textbf{0.00}  & \textbf{0.00}  & 0.06  & 0.02   \\ 
                    &                         & Dense  &  1050  &\textbf{0.00}  & \textbf{0.00}  & \textbf{0.00}  & 0.05  & 0.02   \\ 
\cmidrule{2-9} 
                     &\multirow{4}{*}{\textsf{Cluster}}& Sparse  &  400  &\textbf{0.00}  & \textbf{0.00}  & \textbf{0.00}  & 0.03  & 0.01   \\ 
                    &                         & Sparse  &  1050  &\textbf{0.00}  & \textbf{0.00}  & \textbf{0.00}  & 0.02  & \textbf{0.00}   \\ 
                    &                         & Dense  &  400  &\textbf{0.00}  & \textbf{0.00}  & \textbf{0.00}  & 0.05  & 0.01   \\ 
                    &                         & Dense  &  1050  &\textbf{0.00}  & \textbf{0.00}  & \textbf{0.00}  & 0.04  & 0.01   \\ 
\cmidrule{2-9} 
                  &\multirow{2}{*}{\textsf{Scale-free}}& Sparse  &  400  &\textbf{0.00}  & \textbf{0.00}  & \textbf{0.00}  & 0.03  & 0.01   \\ 
                    &                         & Sparse  &  1050  &\textbf{0.00}  & \textbf{0.00}  & \textbf{0.00}  & 0.02  & 0.01   \\ 
\cmidrule{1-9} 
\multirow{10}{*}{100}&\multirow{4}{*}{\textsf{Random}} & Sparse  &  40  &0.02  & 0.02  & 0.04  & 0.04  & \textbf{0.01}   \\ 
                    &                         & Sparse  &  700  &\textbf{0.00}  & \textbf{0.00}  & 0.01  & 0.01  & 0.01   \\ 
                    &                         & Dense  &  40  &0.02  & 0.02  & 0.05  & 0.05  & \textbf{0.01}   \\ 
                    &                         & Dense  &  700  &\textbf{0.00}  & \textbf{0.00}  & 0.01  & 0.02  & 0.01   \\ 
\cmidrule{2-9} 
                     &\multirow{4}{*}{\textsf{Cluster}}& Sparse  &  40  &0.02  & 0.02  & 0.03  & 0.04  & \textbf{0.01}   \\ 
                    &                         & Sparse  &  700  &\textbf{0.00}  & \textbf{0.00}  & 0.01  & 0.01  & 0.01   \\ 
                    &                         & Dense  &  40  &0.02  & 0.02  & 0.05  & 0.05  & \textbf{0.01}   \\ 
                    &                         & Dense  &  700  &\textbf{0.00}  & \textbf{0.00}  & 0.01  & 0.02  & 0.01   \\ 
\cmidrule{2-9} 
                  &\multirow{2}{*}{\textsf{Scale-free}}& Sparse  &  40  &0.02  & 0.02  & 0.04  & 0.04  & \textbf{0.01}   \\ 
                    &                         & Sparse  &  700  &\textbf{0.00}  & \textbf{0.00}  & 0.01  & 0.01  & 0.01   \\ 
\cmidrule{1-9} 
\multirow{10}{*}{10}  &\multirow{4}{*}{\textsf{Random}}& Sparse  &  20  &0.04  & 0.04  & 0.05  & 0.04  & \textbf{0.01}   \\ 
                    &                         & Sparse  &  350  &\textbf{0.01}  & \textbf{0.01}  & \textbf{0.01}  & \textbf{0.01}  & \textbf{0.01}   \\ 
                    &                         & Dense  &  20  &0.07  & 0.07  & 0.09  & 0.07  & \textbf{0.05}   \\ 
                    &                         & Dense  &  350  &\textbf{0.01}  & \textbf{0.01}  & 0.03  & 0.02  & 0.03   \\ 
\cmidrule{2-9} 
                     &\multirow{4}{*}{\textsf{Cluster}}& Sparse  &  20  &0.05  & 0.05  & 0.05  & 0.04  & \textbf{0.02}   \\ 
                    &                         & Sparse  &  350  &\textbf{0.01}  & \textbf{0.01}  & \textbf{0.01}  & \textbf{0.01}  & \textbf{0.01}   \\ 
                    &                         & Dense  &  20  &0.04  & 0.04  & 0.05  & 0.04  & \textbf{0.00}   \\ 
                    &                         & Dense  &  350  &\textbf{0.00}  & \textbf{0.00}  & 0.01  & 0.02  & 0.01   \\ 
\cmidrule{2-9} 
                  &\multirow{2}{*}{\textsf{Scale-free}}& Sparse  &  20  &0.05  & 0.05  & 0.06  & 0.05  & \textbf{0.02}   \\ 
                    &                         & Sparse  &  350  &\textbf{0.01}  & \textbf{0.01}  & 0.02  & \textbf{0.01}  & 0.03   \\ 
\bottomrule
\end{tabular}
\caption{\textit{$Pr^-$ scores of the algorithms for different instances. The $Pr^-$ reaches its best score at $0$ and its worst at $1$. The values are averages over 16 replications for $p \in \{10, 100\}$ and over 8 replications for $p=1000$. For each setting, the best-performing algorithm is highlighted in \textbf{bold}.}}
\label{table:pmin}
\end{table}
\vspace*{\fill}

\vspace*{\fill}
\begin{table}[ht]
\centering
\begin{tabular}{c @{\hspace{0.7em}} c @{\hspace{0.7em}} c c @{\hspace{1.6em}} c c c @{\hspace{1.7em}} c @{\hspace{0.7em}} c}
\arrayrulecolor{blue}\toprule
$p$   & Graph  & Density & $n$ &  RJ-MPL & BD-MPL & BD & SS & B-CON  \\ 
\hline 
\multirow{10}{*}{1000}&\multirow{4}{*}{\textsf{Random}}& Sparse  &  400  &16000K & 300K & 10 & 600 & 40K  \\ 
                    &                         & Sparse  &  1050  &16000K & 200K & 10 & 400 & 40K  \\ 
                    &                         & Dense  &  400  &30000K & 1500K & 10 & 1500 & 250K  \\ 
                    &                         & Dense  &  1050  &10000K & 500K & 10 & 1500 & 80K  \\ 
\cmidrule{2-9} 
                     &\multirow{4}{*}{\textsf{Cluster}}& Sparse  &  400  &16000K & 300K & 10 & 600 & 40K  \\ 
                    &                         & Sparse  &  1050  &16000K & 200K & 10 & 400 & 40K  \\ 
                    &                         & Dense  &  400  &30000K & 1500K & 10 & 1500 & 250K \\ 
                    &                         & Dense  &  1050  &30000K & 500K & 10 & 1500 & 80K  \\ 
\cmidrule{2-9} 
                  &\multirow{2}{*}{\textsf{Scale-free}}& Sparse  &  400  &30000K & 200K & 10 & 600 & 50K  \\ 
                    &                         & Sparse  &  1050  &30000K & 200K & 10 & 200 & 50K  \\ 
\cmidrule{1-9} 
\multirow{10}{*}{100}&\multirow{4}{*}{\textsf{Random}} & Sparse  &  40  &125000K & 2500K & 30K & 45K & 400K  \\ 
                    &                         & Sparse  &  700  &125000K & 2500K & 30K & 45K & 400K  \\ 
                    &                         & Dense  &  40  &125000K & 2500K & 30K & 45K & 400K  \\ 
                    &                         & Dense  &  700  &125000K & 2500K & 30K & 45K & 400K  \\ 
\cmidrule{2-9} 
                     &\multirow{4}{*}{\textsf{Cluster}}& Sparse  &  40  &125000K & 2500K & 30K & 45K & 400K  \\ 
                    &                         & Sparse  &  700  &125000K & 2500K & 30K & 45K & 400K  \\ 
                    &                         & Dense  &  40  &125000K & 2500K & 30K & 45K & 400K  \\ 
                    &                         & Dense  &  700  &125000K & 2500K & 30K & 45K & 400K  \\ 
\cmidrule{2-9} 
                  &\multirow{2}{*}{\textsf{Scale-free}}& Sparse  &  40  &125000K & 2500K & 30K & 45K & 400K  \\ 
                    &                         & Sparse  &  700  &125000K & 2500K & 30K & 45K & 400K  \\ 
\cmidrule{1-9} 
\multirow{10}{*}{10}  &\multirow{4}{*}{\textsf{Random}}& Sparse  &  20  &100K & 30K & 30K & 3K & 10K  \\ 
                    &                         & Sparse  &  350  &100K & 30K & 30K & 3K & 10K  \\ 
                    &                         & Dense  &  20  &100K & 30K & 30K & 3K & 10K  \\ 
                    &                         & Dense  &  350  &100K & 30K & 30K & 3K & 10K  \\ 
\cmidrule{2-9} 
                     &\multirow{4}{*}{\textsf{Cluster}}& Sparse  &  20  &100K & 30K & 30K & 3K & 10K  \\ 
                    &                         & Sparse  &  350  &100K & 30K & 30K & 3K & 10K  \\ 
                    &                         & Dense  &  20  &100K & 30K & 30K & 3K & 10K  \\ 
                    &                         & Dense  &  350  &100K & 30K & 30K & 3K & 10K  \\ 
\cmidrule{2-9} 
                  &\multirow{2}{*}{\textsf{Scale-free}}& Sparse  &  20  &100K & 30K & 30K & 3K & 10K  \\ 
                    &                         & Sparse  &  350  &100K & 30K & 30K & 3K & 10K  \\ 
\bottomrule
\end{tabular}
\caption{\textit{Number of MCMC iterations until AUC-PR convergence for different instances. The time limit was set to five days, which is why the number of iterations for the BD algorithm for cases with \(p = 1000\) is only 10.}}
\label{table:iterations}
\end{table}
\vspace*{\fill}

\section{Additional Materials for Application to Human Gene Expression}
\label{subsec:mediumdata}

Here, we present additional results from Application to Human Gene Expression, comparing the output of five algorithms using two metrics: the average absolute differences in edge inclusion probabilities for all unique edges, shown in Table \ref{table:difference}, and the percentage of edges identified by method A that are also detected by method B, using a threshold of 0.9 for edge inclusion probability, presented in Table \ref{tab:overlap_0.9}. The average absolute differences in edge inclusion probabilities in Table \ref{table:difference} are relatively low but are influenced by the presence of many edges with inclusion probabilities close to zero.

Table \ref{tab:overlap_0.9} demonstrates that, with a 0.9 threshold for edge inclusion probabilities, B-CON identifies the highest number of edges (87), followed by RJ-MPL and BD-MPL (75 and 73, respectively), BD (68), and SS (35). Starting with SS, which identifies the fewest edges, Table \ref{tab:overlap_0.9} shows that nearly all edges identified by SS are also identified by the other algorithms. For BD, 68\% to 79\% of its identified edges overlap with those identified by other algorithms, such as B-CON, RJ-MPL, and BD-MPL, which have a higher number of identified edges. Notably, B-CON, BD-MPL, and RJ-MPL exhibit substantial overlap: approximately 71\% to 75\% of B-CON's edges are also identified by BD-MPL and RJ-MPL, respectively, while 97\% of BD-MPL's edges overlap with those identified by RJ-MPL.

\begin{table}[ht]
\centering
\arrayrulecolor{blue}
\begin{tabular}{l | @{\hspace{1.5em}} c c c @{\hspace{2em}} c c}
\toprule
          & BD-MPL & RJ-MPL & BD & SS & B-CON \\
\hline 
BD-MPL    &-   & 0.005 & 0.067 & 0.077 & 0.023  \\
RJ-MPL    & -   & -    & 0.067 & 0.077 & 0.023  \\
BD        & -   & -     & -    & 0.045 & 0.074  \\
SS        & -   & -     & -     & -    & 0.085  \\
B-CON & -   & -     & -     & -     & -     \\
\bottomrule
\end{tabular}
\caption{\textit{Average absolute difference in edge inclusion probabilities between algorithms on the human gene data set.}  }
\label{table:difference}
\end{table}

\begin{table}[ht]
\centering
\arrayrulecolor{blue}
\begin{tabular}{l | @{\hspace{1.5em}} c c c @{\hspace{2em}} c c}
\toprule
               & BD-MPL & RJ-MPL & BD & SS & B-CON \\
\hline 
BD-MPL (73)    & -    & 0.97 & 0.63 &  0.45 & 0.85 \\
RJ-MPL (75)    & 0.95 & -    & 0.61 &  0.45 & 0.87 \\
BD (68)        & 0.68 & 0.68 & -    &  0.49 & 0.79 \\
SS (35)        & 0.94 & 0.97 & 0.94 &  -    & 1.00 \\
B-CON (87) & 0.71 & 0.75 & 0.62 & 0.40  & -    \\
\bottomrule
\end{tabular}
\caption{\textit{Proportion of edges identified by the row algorithm that are also found by the column algorithm on the human gene data set, using an edge inclusion probability threshold of 0.9. The numbers in brackets indicate the count of edges with an edge inclusion probability greater than 0.9.}} 
\label{tab:overlap_0.9}
\end{table}

\newpage
\section{Additional Materials for Application to Gene Expression in Immune Cells}
\label{subsec:large-scale data}

Here, we report Tables \ref{table:difference_mice} and \ref{tab:overlap_0.9_mice} for evaluation of the results for Application to Gene Expression in Immune Cells.

\begin{table}[ht]
\centering
\arrayrulecolor{blue}
\begin{tabular}{l | @{\hspace{1.5em}} c c c c}
\toprule
& BD-MPL & RJ-MPL  & SS & B-CON   \\
\hline 
BD-MPL    &-   & 0.019 & 0.026 & 0.071  \\
RJ-MPL    & -   & -    & 0.026 & 0.072   \\
SS        & -   & -     & -    & 0.078  \\
B-CON & -   & -     & -     & -     \\
\bottomrule
\end{tabular}
\caption{\textit{Average absolute difference in edge inclusion probabilities across algorithms for the mice gene data set ($p=623$).}  }
\label{table:difference_mice}
\end{table}

\begin{table}[ht]
\centering
\arrayrulecolor{blue}
\begin{tabular}{l | @{\hspace{1.5em}} c c c c}
\toprule
& BD-MPL & RJ-MPL   & SS  & B-CON \\
\hline 
BD-MPL (3,965)     & - & 0.70    &   0.15  &  0.80 \\
RJ-MPL (4,282)     & 0.65 & -     &   0.14  &  0.78  \\
SS  (656)          & 0.92 & 0.91 & -   &   0.96\\
B-CON (14,258) & 0.22 & 0.23 & 0.04 &  -   \\
\bottomrule
\end{tabular}
\caption{\textit{Proportion of edges identified by the row algorithm that are also found by the column algorithm on the mice data set, using an edge inclusion probability threshold of 0.9. Between brackets is the number of edges with an edge inclusion probability higher than $0.9$.}} 
\label{tab:overlap_0.9_mice}
\end{table}

\begin{figure}[!ht] 
    \centering
    \includegraphics[width=0.6\textwidth]{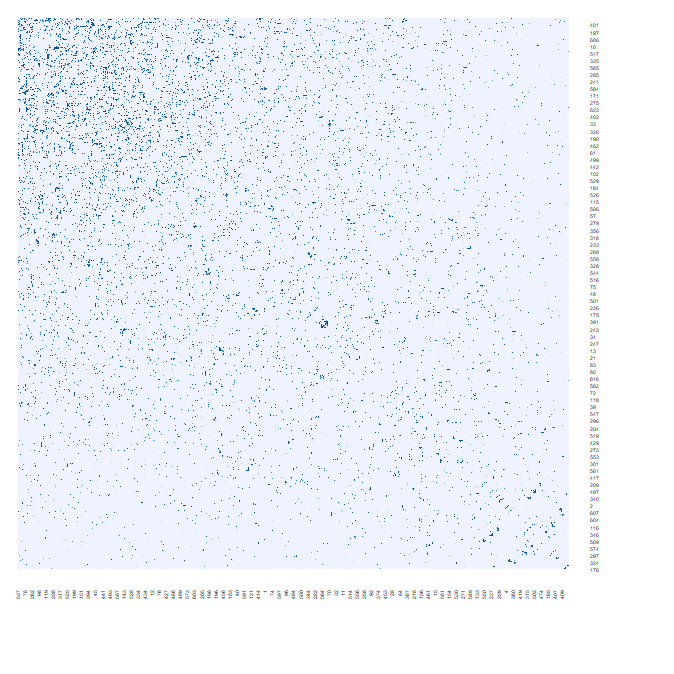}
\caption{\textit{A Heatmap of the edge inclusion probabilities of the BD-MPL algorithm on the mice gene dataset ($p=623$). The probabilities range from $0$ (gray) to $1$ (dark blue).}}
\label{fig:mice_gene_heatmaps}
\end{figure}

\bibliographystyle{Chicago}
\bibliography{ref}

\begin{thebibliography}{}

\bibitem[\protect\citeauthoryear{Albert and Barab{\'a}si}{Albert and
  Barab{\'a}si}{2002}]{albert2002}
Albert, R. and A.-L. Barab{\'a}si (2002).
\newblock Statistical mechanics of complex networks.
\newblock {\em Reviews of Modern Physics\/}~{\em 74\/}(1), 47.

\bibitem[\protect\citeauthoryear{Atay-Kayis and Massam}{Atay-Kayis and
  Massam}{2005}]{atay2005monte}
Atay-Kayis, A. and H.~Massam (2005).
\newblock A {M}onte {C}arlo method for computing the marginal likelihood in
  nondecomposable {G}aussian graphical models.
\newblock {\em Biometrika\/}~{\em 92\/}(2), 317--335.

\bibitem[\protect\citeauthoryear{Atchad{\'e}}{Atchad{\'e}}{2019}]{atchade2019quasi}
Atchad{\'e}, Y.~F. (2019).
\newblock Quasi-bayesian estimation of large gaussian graphical models.
\newblock {\em Journal of Multivariate Analysis\/}~{\em 173}, 656--671.

\bibitem[\protect\citeauthoryear{Avalos-Pacheco, Lazzerini, Lupparelli, and
  Stingo}{Avalos-Pacheco et~al.}{2025}]{avalos2025bayesian}
Avalos-Pacheco, A., A.~Lazzerini, M.~Lupparelli, and F.~C. Stingo (2025).
\newblock Bayesian inference of multiple ising models for heterogeneous public
  opinion survey networks.
\newblock {\em Journal of the Royal Statistical Society Series C: Applied
  Statistics\/}, 1--32.

\bibitem[\protect\citeauthoryear{Barbieri and Berger}{Barbieri and
  Berger}{2004}]{barbieri2004optimal}
Barbieri, M.~M. and J.~O. Berger (2004).
\newblock Optimal predictive model selection.
\newblock {\em Annals of Statistics\/}, 870--897.

\bibitem[\protect\citeauthoryear{Besag}{Besag}{1975}]{besag1975statistical}
Besag, J. (1975).
\newblock Statistical analysis of non-lattice data.
\newblock {\em Journal of the Royal Statistical Society: Series D (The
  Statistician)\/}~{\em 24\/}(3), 179--195.

\bibitem[\protect\citeauthoryear{Bhadra and Mallick}{Bhadra and
  Mallick}{2013}]{Bhadra2013}
Bhadra, A. and B.~Mallick (2013).
\newblock Joint high-dimensional {Bayesian} variable and covariance selection
  with an application to {eQTL} analysis.
\newblock {\em Biometrics\/}~{\em 69}, 447–457.

\bibitem[\protect\citeauthoryear{Cappé, Robert, and Rydén}{Cappé
  et~al.}{2003}]{cappe2003}
Cappé, O., C.~Robert, and T.~Rydén (2003).
\newblock Reversible jump, birth-and-death and more general continuous time
  {M}arkov chain {M}onte {C}arlo samplers.
\newblock {\em Journal of the Royal Statistical Society: Series B (Statistical
  Methodology)\/}~{\em 65\/}(3), 679--700.

\bibitem[\protect\citeauthoryear{Carvalho and Scott}{Carvalho and
  Scott}{2009}]{carvalho2009objective}
Carvalho, C.~M. and J.~G. Scott (2009).
\newblock Objective {B}ayesian model selection in {G}aussian graphical models.
\newblock {\em Biometrika\/}~{\em 96\/}(3), 497--512.

\bibitem[\protect\citeauthoryear{Chandra, M{\"u}ller, and Sarkar}{Chandra
  et~al.}{2024}]{chandra2024bayesian}
Chandra, N.~K., P.~M{\"u}ller, and A.~Sarkar (2024).
\newblock Bayesian scalable precision factor analysis for {Gaussian} graphical
  models.
\newblock {\em Bayesian Analysis\/}~{\em 1\/}(1), 1--29.

\bibitem[\protect\citeauthoryear{Cheng and Lenkoski}{Cheng and
  Lenkoski}{2012}]{cheng2012hierarchical}
Cheng, Y. and A.~Lenkoski (2012).
\newblock Hierarchical {G}aussian graphical models: {B}eyond reversible jump.
\newblock {\em Electronic Journal of Statistics\/}~{\em 6}, 2309--2331.

\bibitem[\protect\citeauthoryear{Colombi, Argiento, Paci, and Pini}{Colombi
  et~al.}{2024}]{colombi2024}
Colombi, A., R.~Argiento, L.~Paci, and A.~Pini (2024).
\newblock Learning block structured graphs in gaussian graphical models.
\newblock {\em Journal of Computational and Graphical Statistics\/}~{\em
  33\/}(1), 152--165.

\bibitem[\protect\citeauthoryear{Consonni and Rocca}{Consonni and
  Rocca}{2012}]{consonni2012objective}
Consonni, G. and L.~L. Rocca (2012).
\newblock Objective {B}ayes factors for {G}aussian directed acyclic graphical
  models.
\newblock {\em Scandinavian Journal of Statistics\/}~{\em 39\/}(4), 743--756.

\bibitem[\protect\citeauthoryear{Davis and Goadrich}{Davis and
  Goadrich}{2006}]{davis2006roc-pr}
Davis, J. and M.~Goadrich (2006).
\newblock The relationship between precision-recall and {ROC} curves.
\newblock In {\em Proceedings of the 23rd International Conference on Machine
  Learning}, New York, pp.\  233--240.

\bibitem[\protect\citeauthoryear{Desch, Randolph, Murphy, Kedl, Lahoud,
  Caminschi, Shortman, Henson, and Jakubzick}{Desch et~al.}{2011}]{Desch2011}
Desch, A., G.~Randolph, K.~Murphy, R.~Kedl, M.~Lahoud, I.~Caminschi,
  K.~Shortman, P.~Henson, and C.~Jakubzick (2011).
\newblock {CD103}+ pulmonary dendritic cells preferentially acquire and present
  apoptotic cell-associated antigen.
\newblock {\em The Journal of Experimental Medicine\/}~{\em 208}, 1789--97.

\bibitem[\protect\citeauthoryear{Dobra, Lenkoski, and Rodriguez}{Dobra
  et~al.}{2011}]{dobra2011bayesian}
Dobra, A., A.~Lenkoski, and A.~Rodriguez (2011).
\newblock Bayesian inference for general {G}aussian graphical models with
  application to multivariate lattice data.
\newblock {\em Journal of the American Statistical Association\/}~{\em
  106\/}(496), 1418--1433.

\bibitem[\protect\citeauthoryear{Dobra and Mohammadi}{Dobra and
  Mohammadi}{2018}]{dobra2015}
Dobra, A. and R.~Mohammadi (2018).
\newblock Loglinear model selection and human mobility.
\newblock {\em The Annals of Applied Statistics\/}~{\em 12\/}(2), 815--845.

\bibitem[\protect\citeauthoryear{Drton and Perlman}{Drton and
  Perlman}{2007}]{drton2007multiple}
Drton, M. and M.~D. Perlman (2007).
\newblock Multiple testing and error control in {G}aussian graphical model
  selection.
\newblock {\em Statistical Science\/}~{\em 22\/}(3), 430--449.

\bibitem[\protect\citeauthoryear{Friedman, Hastie, and Tibshirani}{Friedman
  et~al.}{2008}]{friedman2008sparse}
Friedman, J., T.~Hastie, and R.~Tibshirani (2008).
\newblock Sparse inverse covariance estimation with the graphical {L}asso.
\newblock {\em Biostatistics\/}~{\em 9\/}(3), 432--441.

\bibitem[\protect\citeauthoryear{Gan, Narisetty, and Liang}{Gan
  et~al.}{2019}]{Gan2018}
Gan, L., N.~N. Narisetty, and F.~Liang (2019).
\newblock Bayesian regularization for graphical models with unequal shrinkage.
\newblock {\em Journal of the American Statistical Association\/}~{\em
  114\/}(527), 1218--1231.

\bibitem[\protect\citeauthoryear{Geiger and Heckerman}{Geiger and
  Heckerman}{2002}]{geiger2002parameter}
Geiger, D. and D.~Heckerman (2002).
\newblock Parameter priors for directed acyclic graphical models and the
  characterization of several probability distributions.
\newblock {\em The Annals of Statistics\/}~{\em 30\/}(5), 1412--1440.

\bibitem[\protect\citeauthoryear{Gelman and Rubin}{Gelman and
  Rubin}{1992}]{Gelman1992}
Gelman, A. and D.~B. Rubin (1992).
\newblock Inference from iterative simulation using multiple sequences.
\newblock {\em Statistical Science\/}~{\em 7\/}(4), 457--472.

\bibitem[\protect\citeauthoryear{Green}{Green}{1995}]{green1995reversible}
Green, P. (1995).
\newblock Reversible jump {M}arkov chain {M}onte {C}arlo computation and
  {B}ayesian model determination.
\newblock {\em Biometrika\/}~{\em 82\/}(4), 711--732.

\bibitem[\protect\citeauthoryear{Hanley and Mcneil}{Hanley and
  Mcneil}{1982}]{hanley1982}
Hanley, J. and B.~Mcneil (1982).
\newblock The meaning and use of the area under a receiver operating
  characteristic (roc) curve.
\newblock {\em Radiology\/}~{\em 143\/}(1), 29--36.

\bibitem[\protect\citeauthoryear{Heng, Painter, Elpek, Lukacs-Kornek,
  Mauermann, Turley, Koller, Kim, Wagers, Asinovski, Davis, Fassett, Feuerer,
  Gray, Haxhinasto, Hill, Hyatt, Laplace, Leatherbee, and Kang}{Heng
  et~al.}{2008}]{ImmGen}
Heng, T., M.~Painter, K.~Elpek, V.~Lukacs-Kornek, N.~Mauermann, S.~Turley,
  D.~Koller, F.~Kim, A.~Wagers, N.~Asinovski, S.~Davis, M.~Fassett, M.~Feuerer,
  D.~Gray, S.~Haxhinasto, J.~Hill, G.~Hyatt, C.~Laplace, K.~Leatherbee, and
  J.~Kang (2008).
\newblock The immunological genome project: {N}etworks of gene expression in
  immune cells.
\newblock {\em Nature Immunology\/}~{\em 9}, 1091--1094.

\bibitem[\protect\citeauthoryear{Hinne, Lenkoski, Heskes, and van Gerven}{Hinne
  et~al.}{2014}]{hinne2014efficient}
Hinne, M., A.~Lenkoski, T.~Heskes, and M.~van Gerven (2014).
\newblock Efficient sampling of {G}aussian graphical models using conditional
  {B}ayes factors.
\newblock {\em Stat\/}~{\em 3\/}(1), 326--336.

\bibitem[\protect\citeauthoryear{Jalali, Khare, and Michailidis}{Jalali
  et~al.}{2020}]{jalali2020b}
Jalali, P., K.~Khare, and G.~Michailidis (2020).
\newblock B-{CONCORD}--a scalable {B}ayesian high-dimensional precision matrix
  estimation procedure.
\newblock {\em arXiv preprint arXiv:2005.09017\/}.

\bibitem[\protect\citeauthoryear{Jalali, Khare, and Michailidis}{Jalali
  et~al.}{2023}]{jalali2023bayesian}
Jalali, P., K.~Khare, and G.~Michailidis (2023).
\newblock A {B}ayesian subset specific approach to joint selection of multiple
  graphical models.
\newblock {\em Statistica Sinica\/}~{\em 33}, 1--24.

\bibitem[\protect\citeauthoryear{Koller and Friedman}{Koller and
  Friedman}{2009}]{koller2009probabilistic}
Koller, D. and N.~Friedman (2009).
\newblock {\em Probabilistic graphical models: {P}rinciples and techniques}.
\newblock Cambridge, Massachusetts: MIT Press.

\bibitem[\protect\citeauthoryear{Koschützki and Schreiber}{Koschützki and
  Schreiber}{2008}]{Dirk2008}
Koschützki, D. and F.~Schreiber (2008, 05).
\newblock Centrality analysis methods for biological networks and their
  application to gene regulatory networks.
\newblock {\em Gene Regulation and Systems Biology\/}~{\em 2}, 193--201.

\bibitem[\protect\citeauthoryear{Lauritzen}{Lauritzen}{1996}]{lauritzen1996graphical}
Lauritzen, S.~L. (1996).
\newblock {\em Graphical models}, Volume~17.
\newblock U.K. : Clarendon: Oxford: Oxford University Press.

\bibitem[\protect\citeauthoryear{Leday and Richardson}{Leday and
  Richardson}{2019}]{leday2019fast}
Leday, G.~G. and S.~Richardson (2019).
\newblock Fast bayesian inference in large gaussian graphical models.
\newblock {\em Biometrics\/}~{\em 75\/}(4), 1288--1298.

\bibitem[\protect\citeauthoryear{Lee, Wang, Parisini, Dascher, and
  Nigrovic}{Lee et~al.}{2013}]{lee2013ly6}
Lee, P.~Y., J.-X. Wang, E.~Parisini, C.~C. Dascher, and P.~A. Nigrovic (2013).
\newblock Ly6 family proteins in neutrophil biology.
\newblock {\em Journal of Leukocyte Biology\/}~{\em 94\/}(4), 585--594.

\bibitem[\protect\citeauthoryear{Lenkoski}{Lenkoski}{2013}]{lenkoski2013direct}
Lenkoski, A. (2013).
\newblock A direct sampler for {G}-{W}ishart variates.
\newblock {\em Stat\/}~{\em 2\/}(1), 119--128.

\bibitem[\protect\citeauthoryear{Lenkoski and Dobra}{Lenkoski and
  Dobra}{2011}]{dobra2011computational}
Lenkoski, A. and A.~Dobra (2011).
\newblock Computational aspects related to inference in {G}aussian graphical
  models with the {G}-{W}ishart prior.
\newblock {\em Journal of Computational and Graphical Statistics\/}~{\em 20},
  140--157.

\bibitem[\protect\citeauthoryear{Lepp{\"a}-aho, Johan, Roos, and
  Corander}{Lepp{\"a}-aho et~al.}{2017}]{leppa2017learning}
Lepp{\"a}-aho, J., Johan, T.~Roos, and J.~Corander (2017).
\newblock Learning {G}aussian graphical models with fractional marginal
  pseudo-likelihood.
\newblock {\em International Journal of Approximate Reasoning\/}~{\em 83},
  21--42.

\bibitem[\protect\citeauthoryear{Letac and Massam}{Letac and
  Massam}{2007}]{letac2007whishart}
Letac, G. and H.~Massam (2007).
\newblock Wishart distributions for decomposable graphs.
\newblock {\em The Annals of Statistics\/}~{\em 35\/}(3), 1278--1323.

\bibitem[\protect\citeauthoryear{Li, Craig, and Bhadra}{Li
  et~al.}{2019}]{li2019graphical}
Li, Y., B.~A. Craig, and A.~Bhadra (2019).
\newblock The graphical horseshoe estimator for inverse covariance matrices.
\newblock {\em Journal of Computational and Graphical Statistics\/}~{\em
  28\/}(3), 747--757.

\bibitem[\protect\citeauthoryear{Liang, Buckley, Tu, Langdon, and Tedder}{Liang
  et~al.}{2001}]{liang2001structural}
Liang, Y., T.~R. Buckley, L.~Tu, S.~D. Langdon, and T.~F. Tedder (2001).
\newblock Structural organization of the human ms4a gene cluster on chromosome
  11q12.
\newblock {\em Immunogenetics\/}~{\em 53}, 357--368.

\bibitem[\protect\citeauthoryear{Liu, Lafferty, and Wasserman}{Liu
  et~al.}{2009}]{Liu2009}
Liu, H., J.~Lafferty, and L.~Wasserman (2009).
\newblock The nonparanormal: semiparametric estimation of high dimensional
  undirected graphs.
\newblock {\em Journal of Machine Learning Research\/}~{\em 10\/}(80),
  2295--2328.

\bibitem[\protect\citeauthoryear{Meinshausen and B{\"u}hlmann}{Meinshausen and
  B{\"u}hlmann}{2006}]{meinshausen2006high}
Meinshausen, N. and P.~B{\"u}hlmann (2006).
\newblock High-dimensional graphs and variable selection with the {L}asso.
\newblock {\em The Annals of Statistics\/}~{\em 34\/}(3), 1436--1462.

\bibitem[\protect\citeauthoryear{Mohammadi and Wit}{Mohammadi and
  Wit}{2015}]{mohammadi2015bayesianStructure}
Mohammadi, A. and E.~Wit (2015).
\newblock {B}ayesian structure learning in sparse {G}aussian graphical models.
\newblock {\em Bayesian Analysis\/}~{\em 10\/}(1), 109--138.

\bibitem[\protect\citeauthoryear{Mohammadi}{Mohammadi}{2022}]{mohammadi2022ssgraph}
Mohammadi, R. (2022).
\newblock {\em ssgraph: {B}ayesian graph structure learning using
  spike-and-slab priors}.
\newblock R package version 1.15.

\bibitem[\protect\citeauthoryear{Mohammadi, Massam, and Letac}{Mohammadi
  et~al.}{2023}]{mohammadi2023accelerating}
Mohammadi, R., H.~Massam, and G.~Letac (2023).
\newblock Accelerating {B}ayesian structure learning in sparse {G}aussian
  graphical models.
\newblock {\em Journal of the American Statistical Association\/}~{\em
  118\/}(542), 1345--1358.

\bibitem[\protect\citeauthoryear{Mohammadi, Wit, and Dobra}{Mohammadi
  et~al.}{2024}]{BDgraph}
Mohammadi, R., E.~Wit, and A.~Dobra (2024).
\newblock {\em {BD}graph: {B}ayesian structure learning in graphical models
  using birth-death {MCMC}}.
\newblock R package version 2.73.

\bibitem[\protect\citeauthoryear{Mohammadi and Wit}{Mohammadi and
  Wit}{2019}]{mohammadi2019bdgraph}
Mohammadi, R. and E.~C. Wit (2019).
\newblock {BDgraph}: An {R} package for {B}ayesian structure learning in
  graphical models.
\newblock {\em Journal of Statistical Software\/}~{\em 89\/}(3), 1--30.

\bibitem[\protect\citeauthoryear{Painter, Davis, Hardy, Mathis, and
  Benoist}{Painter et~al.}{2011}]{Painter2011}
Painter, M., S.~Davis, R.~Hardy, D.~Mathis, and C.~Benoist (2011).
\newblock Transcriptomes of the {B} and {T} lineages compared by multiplatform
  microarray profiling.
\newblock {\em The Journal of Immunology\/}~{\em 186\/}(5), 3047--3057.

\bibitem[\protect\citeauthoryear{Pensar, Nyman, Niiranen, and Corander}{Pensar
  et~al.}{2017}]{pensar2016mpl}
Pensar, J., H.~Nyman, J.~Niiranen, and J.~Corander (2017).
\newblock Marginal pseudo-likelihood learning of discrete {M}arkov network
  structures.
\newblock {\em Bayesian Analysis\/}~{\em 12\/}(4), 1195--1215.

\bibitem[\protect\citeauthoryear{Peterson, Stingo, and Vannucci}{Peterson
  et~al.}{2015}]{peterson2015bayesian}
Peterson, C., F.~C. Stingo, and M.~Vannucci (2015).
\newblock Bayesian inference of multiple {G}aussian graphical models.
\newblock {\em Journal of the American Statistical Association\/}~{\em
  110\/}(509), 159--174.

\bibitem[\protect\citeauthoryear{Powers}{Powers}{2020}]{powers2020evaluation}
Powers, D.~M. (2020).
\newblock Evaluation: from precision, recall and {F}-measure to {ROC},
  informedness, markedness and correlation.
\newblock {\em arXiv preprint arXiv:2010.16061\/}.

\bibitem[\protect\citeauthoryear{Preston}{Preston}{1976}]{preston1975}
Preston, C.~J. (1976).
\newblock Special birth-and-death processes.
\newblock {\em Bulletin of the International Statistical Institute\/}~{\em 46},
  371--391.

\bibitem[\protect\citeauthoryear{Raman, Sobolik, and Richmond}{Raman
  et~al.}{2011}]{Raman2011}
Raman, D., T.~Sobolik, and A.~Richmond (2011, 03).
\newblock Chemokines in health and disease.
\newblock {\em Experimental Cell Research\/}~{\em 317}, 575--89.

\bibitem[\protect\citeauthoryear{Roverato}{Roverato}{2002}]{roverato2002hyper}
Roverato, A. (2002).
\newblock Hyper inverse {W}ishart distribution for non-decomposable graphs and
  its application to {B}ayesian inference for {G}aussian graphical models.
\newblock {\em Scandinavian Journal of Statistics\/}~{\em 29\/}(3), 391--411.

\bibitem[\protect\citeauthoryear{Rue and Held}{Rue and
  Held}{2005}]{rue2005gaussian}
Rue, H. and L.~Held (2005).
\newblock {\em Gaussian {M}arkov random fields: {T}heory and applications}.
\newblock London: Chapman and Hall-CRC Press.

\bibitem[\protect\citeauthoryear{Safaee, Clark, Ivan, Oh, Bloch, Sun, Oh, and
  Parsa}{Safaee et~al.}{2013}]{Safaee2013}
Safaee, M., A.~Clark, M.~Ivan, M.~Oh, O.~Bloch, M.~Sun, T.~Oh, and A.~Parsa
  (2013, 08).
\newblock Cd97 is a multifunctional leukocyte receptor with distinct roles in
  human cancers (review).
\newblock {\em International Journal of Oncology\/}~{\em 43}.

\bibitem[\protect\citeauthoryear{Sagar, Banerjee, Datta, and Bhadra}{Sagar
  et~al.}{2024}]{sagar2024precision}
Sagar, K., S.~Banerjee, J.~Datta, and A.~Bhadra (2024).
\newblock Precision matrix estimation under the horseshoe-like prior--penalty
  dual.
\newblock {\em Electronic Journal of Statistics\/}~{\em 18\/}(1), 1--46.

\bibitem[\protect\citeauthoryear{Scutari}{Scutari}{2013}]{scutari2013prior}
Scutari, M. (2013).
\newblock On the prior and posterior distributions used in graphical modelling.
\newblock {\em Bayesian Analysis\/}~{\em 8\/}(3), 505--532.

\bibitem[\protect\citeauthoryear{Stingo and Marchetti}{Stingo and
  Marchetti}{2015}]{stingo2015efficient}
Stingo, F. and G.~M. Marchetti (2015).
\newblock Efficient local updates for undirected graphical models.
\newblock {\em Statistics and Computing\/}~{\em 25}, 159--171.

\bibitem[\protect\citeauthoryear{Stranger, Nica, Forrest, Dimas, Bird, Beazley,
  Ingle, Dunning, Flicek, Koller, Montgomery, Tavaré, Deloukas, and
  Dermitzakis}{Stranger et~al.}{2007}]{Stranger2007}
Stranger, B.~E., A.~C. Nica, M.~S. Forrest, A.~Dimas, C.~P. Bird, C.~Beazley,
  C.~E. Ingle, M.~Dunning, P.~Flicek, D.~Koller, S.~Montgomery, S.~Tavaré,
  P.~Deloukas, and E.~T. Dermitzakis (2007).
\newblock Population genomics of human gene expression.
\newblock {\em Nature Genetics\/}~{\em 39}, 1217--1224.

\bibitem[\protect\citeauthoryear{Tierney}{Tierney}{1994}]{stationary_distr_exists}
Tierney, L. (1994).
\newblock Markov chains for exploring posterior distributions.
\newblock {\em {The Annals of Statistics}\/}~{\em 22\/}(4), 1701 -- 1728.

\bibitem[\protect\citeauthoryear{Uhler, Lenkoski, and Richards}{Uhler
  et~al.}{2018}]{uhler2018exact}
Uhler, C., A.~Lenkoski, and D.~Richards (2018).
\newblock Exact formulas for the normalizing constants of {W}ishart
  distributions for graphical models.
\newblock {\em The Annals of Statistics\/}~{\em 46\/}(1), 90--118.

\bibitem[\protect\citeauthoryear{van~den Boom, Beskos, and De~Iorio}{van~den
  Boom et~al.}{2022}]{van2022g}
van~den Boom, W., A.~Beskos, and M.~De~Iorio (2022).
\newblock The {G}-{W}ishart weighted proposal algorithm: {E}fficient posterior
  computation for {G}aussian graphical models.
\newblock {\em Journal of Computational and Graphical Statistics\/}~{\em
  31\/}(4), 1215--1224.

\bibitem[\protect\citeauthoryear{{v}an~den Boom, De~Iorio, and
  Beskos}{{v}an~den Boom et~al.}{2023}]{vandenboom2023}
{v}an~den Boom, W., M.~De~Iorio, and A.~Beskos (2023).
\newblock Bayesian learning of graph substructures.
\newblock {\em Bayesian Analysis\/}~{\em 18\/}(4), 1311–1339.

\bibitem[\protect\citeauthoryear{Vogels, Mohammadi, Schoonhoven, and
  Birbil}{Vogels et~al.}{2024}]{vogels2024bayesian}
Vogels, L., R.~Mohammadi, M.~Schoonhoven, and {\c{S}}.~{\.I}. Birbil (2024).
\newblock Bayesian structure learning in undirected gaussian graphical models:
  Literature review with empirical comparison.
\newblock {\em Journal of the American Statistical Association\/}~{\em
  119\/}(548), 3164--3182.

\bibitem[\protect\citeauthoryear{Wang}{Wang}{2012}]{wang12}
Wang, H. (2012).
\newblock The {B}ayesian graphical {L}asso and efficient posterior computation.
\newblock {\em Bayesian Analysis\/}~{\em 7}, 771--790.

\bibitem[\protect\citeauthoryear{Wang}{Wang}{2015}]{wang15}
Wang, H. (2015).
\newblock Scaling it up: Stochastic search structure learning in graphical
  models.
\newblock {\em {Bayesian} Analysis\/}~{\em 10\/}(2), 351--377.

\bibitem[\protect\citeauthoryear{Williams and Mulder}{Williams and
  Mulder}{2020}]{williams2020bayesian}
Williams, D.~R. and J.~Mulder (2020).
\newblock Bayesian hypothesis testing for gaussian graphical models:
  Conditional independence and order constraints.
\newblock {\em Journal of Mathematical Psychology\/}~{\em 99}, 102441.

\bibitem[\protect\citeauthoryear{Wolffe}{Wolffe}{2001}]{WOLFFE2001948}
Wolffe, A. (2001).
\newblock Histone genes.
\newblock In S.~Brenner and J.~H. Miller (Eds.), {\em Encyclopedia of
  Genetics}, pp.\  948--952. New York: Academic Press.

\bibitem[\protect\citeauthoryear{Wong, Moffa, and Kuipers}{Wong
  et~al.}{2024}]{wong2024new}
Wong, C., G.~Moffa, and J.~Kuipers (2024).
\newblock A new way to evaluate g-wishart normalising constants via fourier
  analysis.
\newblock {\em arXiv preprint arXiv:2404.06803\/}.

\bibitem[\protect\citeauthoryear{Wong, Moffa, and Kuipers}{Wong
  et~al.}{2025}]{wong2025conjecture}
Wong, C., G.~Moffa, and J.~Kuipers (2025).
\newblock On a conjecture of roverato regarding g-wishart normalising
  constants.
\newblock {\em arXiv preprint arXiv:2503.13046\/}.

\bibitem[\protect\citeauthoryear{Xu, Evans, Bizzaro, Quaglia, Verrillo, Li,
  Stieglmaier, Schiewer, Languino, and Kelly}{Xu et~al.}{2022}]{Xu2022}
Xu, M., L.~Evans, C.~Bizzaro, F.~Quaglia, C.~Verrillo, L.~Li, J.~Stieglmaier,
  M.~Schiewer, L.~Languino, and W.~Kelly (2022, 08).
\newblock Steap1–4 (six-transmembrane epithelial antigen of the prostate
  1–4) and their clinical implications for prostate cancer.
\newblock {\em Cancers\/}~{\em 14}, 4034.

\end{thebibliography}
\end{document}